# On the wave nature of matter:
## A transition from classical mechanics to quantum mechanics

Donald C. Chang

*Macro-Science Group, Division of LIFS, Hong Kong University of Science and Technology, Clear Water Bay, Hong Kong, China*

Email: bochang@ust.hk

**Abstract:**

Following the spirit of de Broglie and Einstein, we think the concepts of matter and radiation can be unified. We know a particle propagates like a wave; its motion is described by certain wave equations. At this point, it is not clear what the wave function represents. Besides the statistical meaning suggested by the Copenhagen interpretation, does the wave function represent any physical motion? For photon, we know it is an electro-magnetic wave. But what about particles with rest mass, such as an electron? To investigate the physical nature of matter wave, we propose that: *(1) Like the photon, a particle is an excitation wave of a real physical field. (2) Different types of particles are different excitation modes of the same field.* Based on this thinking, we show that the concept of quantum mechanics can be a natural extension of classical mechanics. By critically analyzing the transition from classical physics to quantum physics, we found a new physical meaning for the quantum wave function. This work suggests that various quantum wave equations, including the Klein-Gordon equation, the Dirac equation and the Schrödinger equation, could have a common base relating to the Basic Wave Equation of the matter wave. This work has some interesting implications. It suggests a possible way to explain the origin of visible matters and dark matters in our universe.

## 1. Introduction

Quantum mechanics is a great successful theory in physics for the past one hundred years. However, there are still many puzzles about concepts in quantum mechanics today. Unlike the wave equations in classical physics which is based on well-established physical laws, the derivation of quantum wave equations was often based on conjectures; the strongest justification is that these equations can lead to results consistent with experiments [1]-[4]. This is not totally satisfactory. In the tradition of physics, we always want to know the physical basis behind a theory.

In this work, we would like to examine several basic questions related to quantum physics. First, *what is the physical nature of a particle?* It is well known that particles have both properties of wave and corpuscle. This is called "wave-particle duality". In the standard textbooks, this duality is often explained as follows: The particle itself is a pointed object, but its distribution is like a wave. This is called the "Copenhagen interpretation". Such a view, however, is not universally agreed. For example, many well-known physicists, including Schrödinger, de Broglie and Einstein had expressed doubt on such interpretation [5].

Second, *how can one explain the "quantum behavior" of a free particle*? For example, how can a single electron pass two slits simultaneously? How can a single electron be diffracted from a crystal following the Bragg's diffraction law? How can particles be created in the vacuum or disappear into nowhere?

In this work, we would like to explore a hypothesis that regards the sub-atomic particle as an excitation wave of the vacuum. This hypothesis is partially based on the fact that the behavior of an electron is very similar to a photon [6], [7]. From the Planck's relation [8] and Einstein's theory of photoelectric effect [9], we know a photon can behave like a particle. But in reality, we also know that photon is a type of electro-magnetic wave. Thus, we believe that, in the microscopic view, a particle is an excitation wave; but in the macroscopic view, it can behave like a pointed object.

The challenge then is to derive the quantum mechanical wave equations based on the wave hypothesis. This is the main purpose of this work. We will show in the following that the quantum wave equation for the electron can be naturally derived if we assume that the electron is an excitation wave of the vacuum. Our basic approach is to follow the Correspondence Principle, i.e., to regard quantum mechanics as a natural extension of classical mechanics [10]. We will first review the physical basis of mechanical vibration (harmonic oscillator and sound wave in an elastic solid). Then, we will analyze the wave mechanism in an electro-magnetic field (photon). Finally, by extending the classical mechanical treatment of wave propagation, we will develop a model to describe the propagation of matter wave. We will show that this will directly lead us to the various quantum wave equations.

## 2. Wave propagation in a classical mechanical system

To study the physical meaning of the wave function, let us first review what happens in a classical mechanical system. Here, the basic requirement for generating a wave motion involves two types of forces: (a) **Inertial force**, which is related to the change of momentum and thus the kinetic energy; (b) **Restoring force**, which is related to the change of the potential energy. The inertial force is basically described by *Newton's second Law*: $F = ma$. The restoring force is generally described by *Hooke's Law*: $F = -\kappa x$. It is the interaction between these two forces that generates an oscillation in a mechanical system. In the following, we will briefly review three different cases as examples. From these examples, one can see clearly the physical meaning of the wave function.

## 2.1. Wave in a harmonic oscillator

The simplest example of wave generation in a mechanical system is the one-dimensional harmonic oscillator. One can easily set up the equation of motion by equalizing the inertial force with the restoring force. Using Newton's Law, we know

$$F = ma = m\frac{d^2x}{dt^2}. \tag{1}$$

Using Hooke's Law, we have

$$F = -\kappa x. \tag{2}$$

Combining Eqs. (1) and (2), we have the equation of motion for the harmonic oscillator,

$$m\frac{d^2x}{dt^2} = -\kappa x. \tag{3}$$

The most general solution for Eq. (3) is

$$x = x_0 e^{i\omega t}, \tag{4}$$

where $\omega = \sqrt{\kappa/m}$. In this case, the wave function apparently represents the displacement in the harmonic oscillator.

## 2.2. Wave propagation in a one-dimensional string

In the above example, the solution of the wave equation is not a moving wave. Thus, it is not a proper example to demonstrate the generation of a propagating wave. The simplest example to demonstrate wave propagation in a mechanical system is a one-dimensional stretched string.

This string can be modeled as a string of beads, in which the mass of one segment of the string ($\Delta z$) is lumped together to become a bead. (See Figure 1a). Each pair of neighboring beads is then connected by a massless string. The beads can undergo harmonic oscillation. The wave propagating along the string is generated by coupling the harmonic oscillation of neighboring beads.

In such a simplified model, the wave propagating mechanism can be understood very easily. First, the inertial motion of each bead is governed by the Newton's law. Second, the restoring force is governed by the Hooke's law, which is applied to the string connecting two neighboring beads. These two forces interact with each other to allow the wave to travel along the string. In this case, it is more convenient to describe the mechanics using the Lagrangian formulation.

From Newton's Law, we know the kinetic energy of a small segment of the string (length of Δz) is $\Delta T = \frac{1}{2}(\rho \Delta z) v^2$, where $\rho$ is the mass density of the string. Let us denote the vertical displacement of the string as $\phi$, the kinetic energy is

$$\Delta T = \frac{1}{2}(\rho \Delta z)\left(\frac{\partial \phi}{\partial t}\right)^2. \tag{5}$$

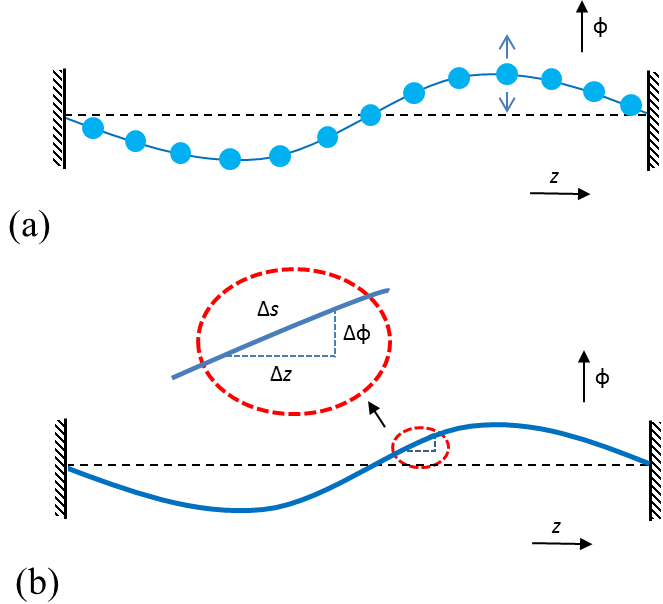

*Figure 1. Wave propagation in a 1-D continuum system (a stretched string). (a) The wave propagation on a string can be modelled as coupled harmonic oscillations of a string of beads. (b) In the more realistic model, each "bead" becomes a tiny segment of the string.*

Using Hooke's Law, one can easily calculate the potential energy of the same segment. As shown in Fig. 1b, the length of the string over $\Delta z$ is stretched to $\Delta s$ during the wave motion. The potential energy for this segment is

$$\Delta V = F_1 (\Delta s - \Delta z)$$

where $F_1$ is the tension of the stretched string between the two end points. From the inset in Fig. 1b, we can see

$$\Delta s = \sqrt{\Delta z^2 + \Delta \phi^2} = \Delta z \left[1 + \frac{\Delta \phi^2}{\Delta z^2}\right]^{\frac{1}{2}} = \Delta z \left[1 + \frac{1}{2}\left(\frac{\Delta \phi}{\Delta z}\right)^2 + \cdots\right]$$

When $\Delta\phi$ is small, we can ignore the higher order terms. So,

$$\Delta V = \frac{F_1}{2}\left(\frac{\partial\phi}{\partial z}\right)^2 \Delta z. \quad (6)$$

From Eqs. (5) and (6), we can obtain the Lagrangian density

$$\mathcal{L} = \frac{1}{2}\rho\left(\frac{\partial\phi}{\partial t}\right)^2 - \frac{1}{2}F_1\left(\frac{\partial\phi}{\partial z}\right)^2. \quad (7)$$

By applying the Euler-Lagrange equation, we can easily obtain the equation of motion

$$\frac{\partial^2\phi}{\partial x^2} - \frac{1}{c_1^2}\frac{\partial^2\phi}{\partial t^2} = 0 \quad (8)$$

where $c_1 = \sqrt{F_1/\rho}$. The general solution for Eq. (8) is

$$\phi = \phi_0 e^{i(z - c_1 t)}. \quad (9)$$

In this case, the wave function apparently represents the transverse displacement of the string.

### 2.3. Wave propagation in an elastic solid

To examine the physical meaning of the wave function in a 3-dimensional mechanical medium, we will consider the wave propagation in an elastic solid. Let us denote the displacement and velocity fields of a differential solid element ($\Delta V$) as $r_i$ and $u_i$ ($i$=1,2,3). From Newton's Law, the time derivative of the momentum (density $\rho \times$ velocity vector $u_i$) is equal to the surface/traction force $T_i$ and the body force $f_i$ applied [11], [12],

$$\frac{d}{dt}(\rho\Delta V)u_i = \rho f_i \Delta V + T_i \Delta S. \quad (10)$$

It is well known that the traction force $T_i$ can be related to the second-order stress tensor $\sigma_{ij}$, i.e., $T_i = \sigma_{ij} n_j$, where $\mathbf{n} = \{n_i\}$ is the normal unit vector with respect to the surface $S$ [11]. It can be shown that [13]

$$T_i \Delta S = \sigma_{ij} n_j \Delta S = \frac{\partial \sigma_{ij}}{\partial x_j} \Delta V. \quad (11)$$

Here we have adopted the Einstein summation convention; repeated indices represent a summation over the 3 axes. Combining Eqs. (10) and (11), we have

$$\frac{d}{dt}(\rho u_i)\Delta V = \frac{\partial \sigma_{ij}}{\partial x_j}\Delta V + \rho f_i \Delta V. \quad (12)$$

Recall that $\frac{d}{dt} = \frac{\partial}{\partial t} + u_k \frac{\partial}{\partial x_k}$, Eq. (12) can be simplified as:

$$\rho\frac{\partial u_i}{\partial t} + \rho u_k \frac{\partial u_i}{\partial x_k} = \frac{\partial \sigma_{ij}}{\partial x_j} + \rho f_i. \tag{13}$$

If the deformation of the elastic body is small, we can neglect the second-order terms. Recall that the velocity vector $u_i$ is the time derivative of the displacement $r_i$, we have

$$\frac{\partial \sigma_{ij}}{\partial x_j} + \rho f_i = \rho \frac{\partial u_i}{\partial t} = \rho \frac{\partial^2 r_i}{\partial t^2}. \tag{14}$$

If the material in the solid is linear, isotropic and the deformation is small, the strain tensor $e_{ij}$ can be related to the average of the deformation gradient $r_{i,j}$ [14],

$$e_{ij} = \frac{1}{2}(r_{i,j} + r_{j,i}), \tag{15}$$

where $r_{i,j} = \frac{\partial r_i}{\partial x_j}$. In addition, the stress tensor for isotropic material is known to be related to the strain tensor according to the generalized Hook's law with only two constants [11], [14]:

$$\sigma_{ij} = \lambda e_{kk} \delta_{ij} + 2\mu e_{ij} = \lambda r_{k,k} \delta_{ij} + \mu(r_{i,j} + r_{j,i}), \tag{16}$$

where $\lambda$ and $\mu$ are *Lamé's first parameter* and *Lamé's second parameter*. $\mu$ is also called the "*shear modulus*"[15]. Substituting Eq. (16) into Eq. (14), we can obtain the following equation (called the "*Navier equation*"), which is the fundamental equation for describing wave propagation in solid mechanics [14]:

$$(\lambda + \mu) r_{j,ji} + \mu r_{i,jj} + \rho f_i = \rho \ddot{r}_i. \tag{17}$$

Since there is no external force applied to the elastic solid, the body force $f_i$ is zero here. Using the vector notation of the gradient and divergence operators, we can re-write the tensor equation of Eq. (17) as the following:

$$(\lambda + \mu)\nabla(\nabla \cdot \mathbf{r}) + \mu \nabla^2 \mathbf{r} = \rho \frac{\partial^2 \mathbf{r}}{\partial t^2}. \tag{18}$$

## 2.4. Application of the Helmholtz decomposition theorem on the wave motion of an elastic solid

Unlike the one-dimensional string, the wave motion in an elastic solid is more complicated, since **r** can take on different oscillating modes. Based on the Helmholtz decomposition theorem [16], the vector **r** can be decomposed into a curl-free component $\phi$ and a divergence-free component $\mathbf{\psi}$:

$$\mathbf{r} = -\nabla \phi + \nabla \times \mathbf{\psi}, \tag{19}$$

where $\nabla \cdot \mathbf{\psi} = 0$. (For more details, see *Appendix*). Substituting Eq. (19) into Eq. (18), we have

$$(\lambda + \mu)\nabla(\nabla \cdot [-\nabla \phi + \nabla \times \mathbf{\psi}]) + \mu \nabla^2 [-\nabla \phi + \nabla \times \mathbf{\psi}] = \rho \frac{\partial^2}{\partial t^2}[-\nabla \phi + \nabla \times \mathbf{\psi}].$$

This equation can be rearranged to become

$$\nabla\left\{(\lambda+2\mu)\nabla^2\phi - \rho\frac{\partial^2\phi}{\partial t^2}\right\} = \nabla\times\left\{\mu\nabla^2\psi - \rho\frac{\partial^2\psi}{\partial t^2}\right\}. \qquad (20)$$

Since $\phi$ and $\psi$ are independent from each other, the simplest way to satisfy Eq. (20) is to assume that each bracketed term is equal to zero. Therefore, Eq. (20) implies that we can have two uncoupled wave equations:

$$\begin{cases} (\lambda+2\mu)\nabla^2\phi - \rho\dfrac{\partial^2\phi}{\partial t^2} = 0 & (21) \\[2mm] \mu\nabla^2\psi - \rho\dfrac{\partial^2\psi}{\partial t^2} = 0. & (22) \end{cases}$$

Re-arranging the coefficients of Eqs. (21) and (22), we have

$$\begin{cases} \nabla^2\phi - \dfrac{1}{c_p^2}\dfrac{\partial^2\phi}{\partial t^2} = 0 & (23) \\[2mm] \nabla^2\psi - \dfrac{1}{c_s^2}\dfrac{\partial^2\psi}{\partial t^2} = 0, & (24) \end{cases}$$

where $c_p = \sqrt{(\lambda+2\mu)/\rho}$ is the velocity of the dilational wave (also called "primary wave") and $c_s = \sqrt{\mu/\rho}$ is the velocity of the transverse wave (also called "distortional wave/secondary wave/shear wave").

These wave equations have a similar form as the vibrating string as shown in Eq. (8). The meanings of the wave functions, however, are slightly different. In the case of the 1-D string, the wave function represents a transverse displacement of the string. But in the case of an elastic solid, the wave functions $\phi$ and $\psi$ do not directly represent the displacement of the solid element. Instead, they represent potential functions, the derivatives of which are related to different modes of displacement according to the Helmholtz decomposition. As will be shown later, what we learn here for the elastic solid can be highly useful when we try to analyze the propagation of matter wave in the vacuum.

## 3. Wave propagation in an electro-magnetic system

Now, let us examine the mechanism of wave propagation in an electro-magnetic system. Following the approach of Maxwell, we will regard the electro-magnetic wave as an excitation of the vacuum. Unlike the mechanical systems, Newton's law and Hooke's law cannot be applied here. First, there is no inertial force, since the vacuum does not have any rest mass. Second, there is no restoring force, since there is no mechanical coupling between the neighboring volume elements of the vacuum. One cannot treat the vacuum as an elastic solid, and thus, it is impossible to set up a wave equation like the Navier equation. Then, the electro-magnetic system must use a different mechanism to generate a propagating wave.

### 3.1. Wave propagation based on the Maxwell's equations

In the electro-magnetic system, the wave regenerating mechanism is really based on the Maxwell's equations, more specifically, the Ampere's law and the Faraday's law. Let us recall that the Maxwell's equations include:

$$\begin{cases} \nabla \times \mathbf{E} = -\dfrac{\partial \mathbf{B}}{\partial t} & \text{Faraday's Law} \quad (25) \\[6pt] \nabla \times \mathbf{H} = \mathbf{J} + \dfrac{\partial \mathbf{D}}{\partial t} & \text{Ampère's Law} \quad (26) \\[6pt] \nabla \cdot \mathbf{D} = \rho_e & \text{Coulomb's Law (or Gauss's Law)} \quad (27) \\[6pt] \nabla \cdot \mathbf{B} = 0 & \text{Gauss' Law for Magnetism} \quad (28) \end{cases}$$

In the above, **E** is the electric field and **H** is the magnetic field; **B** and **D** are magnetic induction and electric displacement, respectively; and

$$\begin{cases} \mathbf{B} = \mu \mathbf{H} & (29) \\ \mathbf{D} = \varepsilon \mathbf{E}, & (30) \end{cases}$$

where $\varepsilon$ and $\mu$ are the dielectric permittivity and magnetic permeability. In the vacuum, the external current $\mathbf{J} = 0$, and $\rho_e = 0$; also $\mu = \mu_0$, $\varepsilon = \varepsilon_0$. We can re-write the first two Maxwell's equations as

$$\begin{cases} \nabla \times \mathbf{E} = -\mu_0 \dfrac{\partial \mathbf{H}}{\partial t} & (25A) \\[6pt] \nabla \times \mathbf{H} = \varepsilon_0 \dfrac{\partial \mathbf{E}}{\partial t}\ . & (26A) \end{cases}$$

One can easily see that **E** and **H** can cross interact with each other in the vacuum. When **E** changes with time, it generate a change in curl **H** in the vicinity; and when **H** changes with time, it would induce **E** around it to change. This cross interaction allows the electro-magnetic wave to propagate in the vacuum.

Hence, one can easily see that, there is a close analogy between the wave-generating mechanism in the mechanical system and that in the electro-magnetic system. In the mechanical system, it is the interactions between the general coordinate (*q*) and the general momentum (*p*) as described in Newton's Law and Hooke's Law that generates an oscillating wave. In the electro-magnetic system, however, it is the cross interactions between **E** and **H** that generates a propagating wave. This comparison can be clearly seen in the following table:

*Table 1*. Corresponding relations in mechanical system and electro-magnetic system

| Mechanical system (Newton's Law & Hooke's Law) | Electro-magnetic system (Maxwell's equations: Faraday's law and Ampere's law) |
|---|---|
| $\dfrac{dp}{dt} = -kq$ | $\dfrac{\partial \mathbf{H}}{\partial t} = -\dfrac{1}{\mu} \nabla \times \mathbf{E}$ |
| $\dfrac{dq}{dt} = \dfrac{1}{m} p$ | $\dfrac{\partial \mathbf{E}}{\partial t} = \dfrac{1}{\varepsilon} \nabla \times \mathbf{H}$ |

In these two systems, $q$ and $p$, $\mathbf{H}$ and $\mathbf{E}$ are both related with the time derivatives of their counter parts. The only difference is that, in the mechanical system, the variations of $q$ and $p$ are directly proportional to $dp/dt$ and $dq/dt$; but in the electro-magnetic system, $\mathbf{H}$ and $\mathbf{E}$ are not directly proportional to the time derivatives. Instead, the curls of $\mathbf{E}$ and $\mathbf{H}$ are proportional to the time derivatives of their counter parts. This shows a fundamental difference of the two systems.

### 3.2. Wave equation for electro-magnetic radiation (photon)

Once we know the foundation of the wave generating mechanism, it is not difficult to set up the wave equation. From Eqs. (25A) and (26A), one can easily derive the wave equation for an electro-magnetic wave.

Take the curl operation on both side of Eq. (25A),

$$\nabla \times (\nabla \times \mathbf{E}) = -\frac{\partial}{\partial t}(\nabla \times \mathbf{B}).$$

Recall that $\nabla \times (\nabla \times \mathbf{E}) = \nabla(\nabla \cdot \mathbf{E}) - \nabla^2 \mathbf{E}$, and applying Eqs. (26A) and (29),

$$\nabla(\nabla \cdot \mathbf{E}) - \nabla^2 \mathbf{E} = -\mu_0 \varepsilon_0 \frac{\partial}{\partial t}\left(\frac{\partial \mathbf{E}}{\partial t}\right).$$

Since in the vacuum, $\rho_e = 0$, so $\nabla \cdot \mathbf{E} = 0$, the above equation becomes

$$\nabla^2 \mathbf{E} - \frac{1}{c^2} \frac{\partial^2 \mathbf{E}}{\partial t^2} = 0, \tag{31}$$

where $c = 1/\sqrt{\mu_0 \varepsilon_0}$. Using similar operations on Eq. (26A), one can also derive

$$\nabla^2 \mathbf{H} - \frac{1}{c^2} \frac{\partial^2 \mathbf{H}}{\partial t^2} = 0.$$

Thus, the cross interactions between the electric field **E** and magnetic field **H** can generate a propagating wave in the electro-magnetic system.

## 4. What is the physical nature of matter wave?

So far, we have reviewed the wave propagation mechanisms in the mechanical system and the electro-magnetic system. Both the mechanical waves and the electro-magnetic radiation waves (photons) are clearly physical waves, i.e., their wave function represents the movement of physical parameters. Furthermore, these waves are carried by their corresponding media. In quantum mechanics, particles with mass are known to exhibit some wave properties. They are called "matter waves" [17]. The questions now are that: Is this matter wave a physical wave? Is it carried by a medium? And, if yes, what is this medium?

**4.1. Evidence suggesting that the electron is a physical wave**

In the traditional interpretation of the wave-particle duality (the "Copenhagen interpretation"), the particle is assumed to be a pointed object (like a tiny bullet), but its distribution is like a wave [5], [18]. According to this view, the wave function only has a statistical meaning. That is, the magnitude square of the wave function gives the probability of finding the presence of a particle at a specific location and time. From the following discussions, one can see that this view may not be totally correct. Although the wave function can give the probability of detecting the particle in experiment; we believe the matter wave is a physical wave and the particle is a wave packet instead of a point-like object. Our reasons are the following:

(1) At present, at least we know one elementary particle (the photon) is a wave. From the Maxwell theory, we know the photon is an oscillation of the electro-magnetic field. So, it is a physical wave. Furthermore, the size of the photon apparently cannot be regarded as a "point-like" object. For visible light, their wavelength is about 400- 700 nm. The wave packet making up the photon must span over many wavelengths, it is clearly not a point-like object.

(2) Is the electron a point-like object or a wave? It has been demonstrated in many experiments that an electron behaves just like a photon [6]. For example, it was shown a long time ago that electrons can be diffracted by a crystal following the Bragg's law [19], [20]. If the electron is a point-like object, it can only bounce from one atom in the crystal, and thus should not form an interference pattern following the Bragg's law. Furthermore, it has been shown experimentally that a single electron can pass through a double-slit to give an interference pattern similar to that of light [21]. How can a point-like electron do that? If the electron is really a corpuscular object, it can only pass one of the slits at one time, and thus, it could never form an interference pattern with itself passing through the other slit. So far, there is no satisfactory explanation on how a "probability wave" of a pointed object can form an interference pattern.

(3) Furthermore, if the electron is a point-like object, the atom will be vastly empty. How can an atom behave like a hard sphere? How can the point-like electrons hold atoms together to form molecule? Can a "probability wave" form chemical bonds between atoms and hold them together?

The only logical explanation is that the electron is a *wave* in nature. It should be a *physical wave* instead of probability wave. In fact, it is because the electron is a physical wave that allows us to build electron microscopes [22]. If the electron is a corpuscular object, how can a transmitting electron microscope work?

Another strong reason for us to believe that the electron is a wave is the fact that an electron can be created or annihilated in the vacuum. It has been demonstrated that an electron-positron pair can be created by energetic photon. It is very difficult to explain these observations if one regards the electron as a massive pointed object. On the other hand, such observations can be explained very easily if one believes that electrons (and positrons) are excitation waves of the vacuum, since new waves can be generated by an excitation.

One may say that the creation/annihilation of electrons was already explained by the Dirac theory, which assumed that there is an infinite "sea" of "negative energy electrons" in the vacuum [23], [24]. When the vacuum is excited by an energetic photon, it causes one of the electrons in the negative energy state to become unoccupied. This creates a hole in the negative energy state. This hole would behave like an anti-particle of the electron (i.e., the positron). Although the Dirac theory is widely used, there are still many problems with it. First, it has very complicated assumptions. As far as we know, there is no physical evidence supporting the idea that the vacuum is filled with a sea of negative energy electrons. Second, it has troublesome implications. Besides electrons, many particles can be created or annihilated in the vacuum. If one uses the Dirac theory to explain their creation and annihilation, it would require the existence of many "seas" of particles occupying all negative energy states. Thus, the vacuum is not only full of negative energy electrons; it is also filled with infinite numbers of negative energy muons, neutrinos, quarks, etc. The vacuum will become a very crowded place.

Finally, the Dirac theory can only be used to explain the creation/annihilation of fermions. It cannot explain why bosons can also be created/annihilated. Since the bosons do not obey Pauli's exclusion principle, one cannot use the Dirac theory to explain hole-creation in the negative energy states.

Hence, the conventional quantum theories based on the particle view still have difficulties to explain many experimental facts. Their physical basis was also not very clear. By comparison, the conceptual picture of the wave model is far simpler. If we regard all particles (including particles and anti-particles with different masses) as different excitation waves of the vacuum, there is no need to assume the existence of an infinite number of particles occupying the negative energy states.

**4.2. Can the vacuum be a wave medium?**

In the history of physics, there were many debates about whether the vacuum is an empty space or not [25]. In Newtonian mechanics, the vacuum is regarded as emptiness, since an object free of applied force can move in straight line at a constant speed. There is nothing in the vacuum that can impede such a motion. But at later time, with the discovery of the electro-magnetic field, many physicists started to assume that there must be a medium that carries the electro-magnetic radiation. This medium is called "aether", which is supposed to occupy all space between matters [25]. This aether hypothesis, however, had many serious problems. The most severe one was its inconsistency with experimental observations. In late 19[th] century, many scientists attempted to use optical interfereometers to detect the movement between aether and the Earth. The most famous one was the Michelson-Morley experiment [26]. None of such measurements was able to detect any movement between Earth and the hypothetical aether. Furthermore, the aether hypothesis was later thought to be unnecessary. In 1905, Einstein proposed the Special Theory of Relativity (STR) and showed that one can explain the null results easily without the assumption of aether [27]. So, this aether hypothesis was totally abandoned in the early 20[th] century.

In most physics textbooks today, the vacuum is regarded as an empty space. This view, however, does not appear to be consistent with the modern theories in cosmology and particle physics. In the Standard Model of cosmology today, the vacuum has very complicated features; it is just the ground state of the cosmos. The energy of our existing universe is supposed to come from the quantum fluctuation in the vacuum [28]. An empty vacuum is also not consistent with the current theories of quantum physics. For example, in quantum electrodynamics, every oscillation mode of radiation is supposed to have a zero-point energy [29]. Such energy is assumed to be a part of the vacuum system. In fact, in the quantum field theory, the vacuum is always regarded as the ground state. The physical fields are just excitations above the vacuum [30].

Today, we can no longer treat the vacuum as emptiness, although its physical properties are still not well understood. In this work, we will simply regard the vacuum as a medium, the excitation wave of which will appear as elementary particles. Thus, not only radiation wave (photons) is the excitation of the vacuum, matter waves (such as electrons) are also excitation waves of the same vacuum.

We would like to emphasize that the assumption of a vacuum medium is not a revival of the previous aether hypothesis. There are fundamental differences between these two hypotheses. First, the aether was supposed to be a medium filling all space between matters. The vacuum medium, on the other hand, is a pre-existing medium in our universe and it fills all space. Second, the hypothetical aether was assumed to be a medium only to carry the electro-magnetic radiation waves. In our model, both matter waves and radiation waves are excitation waves of the vacuum medium [31]. In another word, all particles found in nature (with or without mass) are excitation waves of the same vacuum medium. As shown in our earlier paper, this vacuum medium model can avoid previous problems encountered by the aether hypothesis [32].

**4.3. The vacuum is a dielectric medium according to Maxwell**

In fact, there is strong evidence suggesting that the vacuum is a medium. In the Maxwell theory of light propagation, it is essential to treat the vacuum as a dielectric medium. An important contribution of Maxwell is his proposal of introducing an "electric displacement" (**D**) in the electro-magnetic theory. In the early version of the Maxwell equations published in 1862, the Ampere's Law was originally written as [33]

$$\nabla \times \mathbf{H} = \mathbf{J} \ . \tag{32}$$

There was a problem in it, because if one performs a divergent operation on the above equation, one will get

$$\nabla \cdot \mathbf{J} = 0.$$

This would be inconsistent with the condition of conservation of charge,

$$\nabla \cdot \mathbf{J} = -\frac{\partial \rho_e}{\partial t} \ . \tag{33}$$

In order to fix this problem, Maxwell proposed to add a new term $\partial \mathbf{D}/\partial t$ into the right hand-side of Eq. (32) [33]. (Here, **D** is called the "electric displacement".) The justification for this was that, for a dielectric material, there are both positive and negative charges embedded within it. When the material is exposed to an electric field, there will be a displacement of the dielectric charges. The time differential of these displaced charges would produce a displacement current ($\mathbf{J}_d$). Maxwell argued that, both the external current **J** and the displacement current $\mathbf{J}_d$ can contribute to the magnetic field. Therefore, Maxwell proposed to reformulate the Ampere's law as

$$\nabla \times \mathbf{H} = \mathbf{J} + \frac{\partial \mathbf{D}}{\partial t} \ . \tag{26}$$

This later became part of the final form of the Maxwell's equations.

The proposal of a displacement current was a major accomplishment by Maxwell. His updated Ampere's Law is widely used today. Furthermore, this updated equation allowed Maxwell to construct his theory of light propagation. When one studies the propagation of electro-magnetic waves in the vacuum, the external current **J** is of course equal to zero. But, **the displacement current $\mathbf{J}_d$ is not assumed to be zero in the vacuum**. This means that, one must regard the vacuum as a dielectric medium. Only with such an assumption that one can justify

$$\nabla \times \mathbf{H} = \frac{\partial \mathbf{D}}{\partial t} \ . \tag{26B}$$

If one regards the vacuum as an empty space, then **D** must automatically equals to zero. It will then be impossible to obtain the correct wave equation of light (i.e., Eq. (31)).

## 5. What is the physical meaning of the wave function?

Based on the above discussions, we can formally hypothesize that: *Matter wave and radiation*

*wave are both excitation waves of the vacuum. Different types of free particles are different excitation modes of this medium.* If this is the case, the wave function of a free particle must represent some sort of motion in the vacuum medium. But what exactly is it?

## 5.1. The wave function should represent the motion of a *field*

To answer this question, we need to review the transition from classical mechanics to quantum mechanics. From our earlier discussions on the mechanical systems, one can see that a propagating wave is an excitation wave of a medium. For example, in the 1D string, the excitation wave function $\phi$ represents the transverse displacement of the string at a local point. The medium in this case is the string. In the case of elastic solid, the wave function represents either the curl-free wave component $\phi$ or the divergence-free wave component $\boldsymbol{\psi}$ of the displacement field. The medium here, of course, is the elastic solid.

From these examples, one can easily see that *the wave function represents the movement of a field*. One should notice that this *field* is not the "*classical field*" (such as the gravitational field or EM field). It is called the "*basic field*". Take the 1-D string as an example, the wave function $\phi$ represents the amplitude of a transverse displacement of the string. So, in such a case, the basic field is a local strain.

In a 3-D system, such as an elastic solid, the *wave function* does not directly represent the displacement itself, but represents the potential functions which indirectly describe the displacement according to the Helmholtz decomposition,

$$\mathbf{r} = -\nabla\phi + \nabla\times\boldsymbol{\psi}.$$

The wave function is either $\phi$ or $\boldsymbol{\psi}$. This suggests that the basic field is the potential functions.

## 5.2. For excitation in the vacuum medium, what does the wave function represent?

Then, one may ask: Is there a direct analogy between the electro-magnetic system and the classical mechanical system?

The answer is yes. In Sections 3 and 4, we have shown that the electro-magnetic radiation wave is an excitation wave of the vacuum medium. The question then is: What is the *basic field* for the electro-magnetic wave? One may think the basic field is either the **E** or **H** fields, since in classical electrodynamics, one can show that both **E** and **H** can satisfy the wave equations

$$\nabla^2\mathbf{E} - \frac{1}{c^2}\frac{\partial^2\mathbf{E}}{\partial t^2} = 0 \quad \text{or} \quad \nabla^2\mathbf{H} - \frac{1}{c^2}\frac{\partial^2\mathbf{H}}{\partial t^2} = 0.$$

But this thinking is actually not correct. Because the Lagrangian density of the electro-magnetic field is [34]

$$\mathcal{L} = \frac{1}{2}\left(\varepsilon\mathbf{E}^2 - \mu\mathbf{H}^2\right). \tag{34}$$

This suggests that neither **E** nor **H** is suitable to play the role of a basic field. Why? This can be seen by comparing the Lagrangian density of the EM system with a mechanical system. Recall that for a 1-D string, the Lagrangian density is

$$\mathcal{L} = \frac{1}{2}\rho\left(\frac{\partial \phi}{\partial t}\right)^2 - \frac{1}{2}F_1\left(\frac{\partial \phi}{\partial z}\right)^2. \tag{7}$$

It shows that *the Lagrangian density should be composed of the quadratic terms of the first derivatives of the basic field*. In comparison with Eq. (7), it is clear that **E** and **H** do not satisfy this requirement.

Taking the hint from the wave propagation in the elastic solid, one may guess that it could be the potential functions of **E** and **H** that play the role of the basic field. According to the Maxwell's theory, **E** and **H** can be derived from the scalar potential Φ and the vector potential **A**. From the Maxwell equations, it is well known that one can express the electric and magnetic fields in Φ and **A** [35]:

$$\begin{cases} \mathbf{B} = \nabla \times \mathbf{A} & (35) \\ \mathbf{E} = -\nabla\Phi - \dfrac{\partial \mathbf{A}}{\partial t}. & (36) \end{cases}$$

In the vacuum, the free charge density $\rho_e = 0$ and thus one can set $\nabla\Phi = 0$. Eq. (36) becomes

$$\mathbf{E} = -\frac{\partial \mathbf{A}}{\partial t}. \tag{36A}$$

Substituting Eqs. (35) and (36A) into Eq. (26A), and using the Coulomb gauge condition $\nabla \cdot \mathbf{A} = 0$, one can easily derive the wave equation

$$\nabla^2 \mathbf{A} - \frac{1}{c^2}\frac{\partial^2 \mathbf{A}}{\partial t^2} = 0, \tag{37}$$

where $c = 1/\sqrt{\mu_0 \varepsilon_0}$ is the speed of light. Eq. (37) suggests that the wave function of the E/M radiation can be the vector potential **A**. This in fact makes good sense. If we express the Lagrangian density of an electro-magnetic field in terms of **A**, it can be shown that it will have a similar form as the Lagrangian density of a mechanical system as shown in Eq. (7). Let us choose a simple system in which the wave is traveling along the *z* axis and the vector potential is along the *x* axis (see Figure 2), i.e., $\mathbf{A} = A_x \hat{x}$. Eqs. (35) and (36A) become,

$$\begin{cases} \mathbf{B} = \left(\dfrac{\partial A_z}{\partial y} - \dfrac{\partial A_y}{\partial z}\right)\hat{x} + \left(\dfrac{\partial A_x}{\partial z} - \dfrac{\partial A_z}{\partial x}\right)\hat{y} + \left(\dfrac{\partial A_y}{\partial x} - \dfrac{\partial A_x}{\partial y}\right)\hat{z} = \dfrac{\partial A_x}{\partial z}\hat{y} & (35A) \\ \mathbf{E} = -\dfrac{\partial \mathbf{A}}{\partial t} = -\dfrac{\partial A_x}{\partial t}\hat{x}. & (35B) \end{cases}$$

Substituting the above into Eq. (34), we have

$$\mathcal{L} = \frac{1}{2}\left[\varepsilon\left|\frac{\partial A_x}{\partial t}\right|^2 - \frac{1}{\mu}\left|\frac{\partial A_x}{\partial z}\right|^2\right] \tag{38}$$

By comparing this with the Lagrangian density of a 1-D string (Eq. (7)), it appears that the vector potential **A** is the counter part of the field parameter $\phi$ in the mechanical system. This suggests that for the photon, its basic field described by the wave function is not the electric field or magnetic field; it is more likely to be its vector potential.

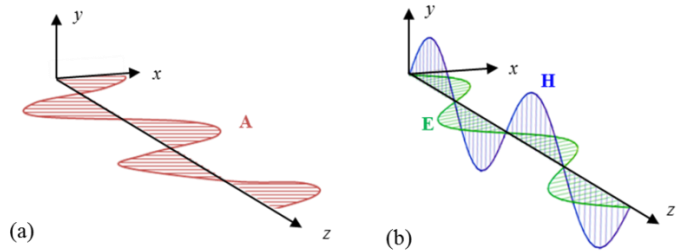

*Figure 2*. ***Electro-magnetic radiation in linearly polarized light***. *(a) Vector potential A oscillates along the x-axis and the wave is traveling along the z-axis. (b) The electric field E oscillates in the x direction; the magnetic field H is perpendicular to E and oscillates in the y direction.*

### 5.3. What is the basic field for the matter wave?

The above analysis provides a very useful hint for us to identify the basic field of the matter wave. In our model, we hypothesize that both the radiation wave and matter wave are excitations of the vacuum medium. Thus, the wave function of the radiation wave and the wave function of matter wave must have similar physical properties. In the case of radiation wave, we know that the basic field described by the wave function is a vector potential of the electro-magnetic field. Then, it is very likely that the wave function of a matter wave must also represent some sort of vector potential function, which is related to the motion of the vacuum medium.

But what is this potential function? Is it the vector potential **A**? Since the quadratic form of the derivatives of **A** can be used to construct a Lagrangian density (see Eq. (38)), one may think the basic field is **A**. We, however, do not think **A** is the best choice for the basic field of the matter wave. The reason is that, if we believe that the matter wave is an excitation wave of the vacuum medium, its wave function should be closely related to the displacement of this medium, just like what happens in the propagation of mechanical wave in an elastic solid. In the latter case, we already know the wave function really describes the potential function ($\phi$ or $\psi$) related to the displacement of the medium (according to the Helmholtz decomposition). This suggests that *the wave function of the matter wave most likely represents a potential function which is related to the displacement of the vacuum.*

Based on such a consideration, the vector potential **A** does not seem to be the best choice. Recall that for an elastic solid, one can use Helmholtz decomposition to separate the longitudinal wave from the transverse wave. (See Section 2). The displacement can be decomposed as: $\mathbf{r} = -\nabla\phi + \nabla\times\mathbf{\psi}$. In an electro-magnetic field, the vector potential **A** is known to be related to the magnetic flux **B**:

$$\mathbf{B} = \nabla\times\mathbf{A}. \tag{35}$$

If **A** is a transverse potential similar to **ψ**, then **B** must be playing the role of a displacement of the vacuum medium (the counter part of **r**). But this can be true only if the vacuum is filled with magnetic charges, so that **B** can represent some sort of magnetic charge displacement. This is highly unlikely, since there is no proof for the existence of magnetic monopoles in our universe. It is difficult to believe that the vacuum is composed of magnetic charges. On the other hand, it is highly reasonable to assume that the vacuum medium is composed of electric charges. First, we know there are electric monopoles in our universe. Second, as we have discussed earlier, the Maxwell theory of EM radiation requires the vacuum to behave as a dielectric medium (see Section 4.3.). In such a case, it is very natural to regard the electric displacement **D** as representing the displacement of the vacuum medium. In another word, **D** in the EM system is the counter part of **r** in an elastic solid. Thus, we may apply the Helmholtz decomposition theorem to decompose **D** into a curl-free component and a divergence-free component,

$$\mathbf{D} = -\nabla\varphi + \nabla\times\mathbf{Z}, \tag{39}$$

where $\nabla\cdot\mathbf{Z} = 0$. We may call **Z** as the "electric vector potential". (From now on, **A** will be referred to as the "magnetic vector potential"). In the vacuum, there is no free charge, $\rho_e = 0$. Thus, $\nabla\cdot\mathbf{D} = -\nabla^2\varphi = 0$. This can be satisfied by choosing $\nabla\varphi = 0$. Eq. (39) then becomes

$$\mathbf{D} = \nabla\times\mathbf{Z}. \tag{39A}$$

Then, the variation of **Z** can indirectly represent the motion of the electric displacement **D**.

### 5.4. Cross interaction between Z and A is responsible for wave excitation in the vacuum

Based on the Maxwell equations, one can show **Z** and **A** can cross interact with each other to generate propagating waves in the vacuum. As we have shown in Section 3.1, there is a symmetrical cross-interaction between **E** and **H**. It is such a cross-interaction that is responsible for generating the electro-magnetic radiation waves. In the following, we will show that **Z** and **A** can also cross-interact with each other, just like **E** and **H**. Such cross-interactions explain why different excitation waves can be generated in the vacuum medium.

From Eqs. (25), (39A), (35), and (30)**,** we have

$$-\frac{\partial\mathbf{B}}{\partial t} = \nabla\times\mathbf{E} = \frac{1}{\varepsilon_0}(\nabla\times\nabla\times\mathbf{Z}) = -\frac{\partial(\nabla\times\mathbf{A})}{\partial t}.$$

This implies

$$\nabla \times \mathbf{Z} = -\varepsilon_0 \frac{\partial \mathbf{A}}{\partial t}. \tag{40}$$

From Eqs. (26A) and (39A), we have

$$\nabla \times \mathbf{H} = \frac{\partial \mathbf{D}}{\partial t} = \frac{\partial (\nabla \times \mathbf{Z})}{\partial t} = \nabla \times \frac{\partial \mathbf{Z}}{\partial t}.$$

Thus,

$$\mathbf{H} = \frac{\partial \mathbf{Z}}{\partial t}. \tag{41}$$

We know $\mathbf{H} = \frac{1}{\mu_0} \mathbf{B} = \frac{1}{\mu_0} (\nabla \times \mathbf{A})$, the above equation becomes

$$\nabla \times \mathbf{A} = \mu_0 \frac{\partial \mathbf{Z}}{\partial t}. \tag{42}$$

By comparing Eqs. (40) and (42), it is clear that the vector potentials $\mathbf{Z}$ and $\mathbf{A}$ can cross-interact with each other. Such cross-interactions will make it possible to generate transverse waves described by $\mathbf{Z}$. For instance, take the curl on both side of Eq. (40), and combine the result with Eq. (42), we have

$$\nabla \times (\nabla \times \mathbf{Z}) = -\varepsilon_0 \frac{\partial (\nabla \times \mathbf{A})}{\partial t} = -\varepsilon_0 \mu_0 \frac{\partial^2 \mathbf{Z}}{\partial t^2}. \tag{43}$$

Since $\nabla \times (\nabla \times \mathbf{Z}) = \nabla (\nabla \cdot \mathbf{Z}) - \nabla^2 \mathbf{Z}$ and $\nabla \cdot \mathbf{Z} = 0$, Eq. (43) becomes

$$\nabla^2 \mathbf{Z} - \frac{1}{c^2} \frac{\partial^2 \mathbf{Z}}{\partial t^2} = 0, \tag{43A}$$

where $c = \sqrt{1/\varepsilon_0 \mu_0}$ is the speed of light. Thus, the excitation wave of the vacuum can be carried by $\mathbf{Z}$. Since we believe the matter wave is an excitation wave of the vacuum, and the *basic field* should represent a movement of the medium, it appears that $\mathbf{Z}$ would satisfy both requirements. Thus, we may identify *the basic field for the matter wave as the divergence-free component ($\mathbf{Z}$) of the electric displacement $\mathbf{D}$*.

### 5.5. There is a close analogy between wave mechanisms in the mechanical system and the vacuum medium

From the above discussions, one can see that there is a close analogy between the wave generating mechanisms in a mechanical medium and those in the vacuum. Table 2 is a detailed comparison between different types of wave generating mechanisms. When a mechanical medium (such as an elastic solid) is undergoing an excitation, its strain is represented by a displacement vector $\mathbf{r}$, and its stress (local force/field) is a tensor related to the strain via the generalized Hooke's law. Similarly, when the vacuum medium is undergoing an excitation, it produces an electric displacement **D**; the local electric field **E** is related to **D** through a relation analogous to Hooke's law, $\mathbf{E} = \frac{1}{\varepsilon_0}\mathbf{D}$. Furthermore, the displacement vector **r** can be decomposed into a curl-free component $\phi$ and a divergence-free component $\psi$. Similarly, the electric displacement vector **D** can also be decomposed into a curl-free component $\nabla\varphi$ and a divergence-free component $\nabla\times\mathbf{Z}$. The only difference is that here $\nabla\varphi = 0$ in the vacuum. This means that, unlike the elastic solid, the vacuum medium has no longitudinal wave; it can only generate transverse waves.

*Table 2. Comparison of wave generating mechanisms between the mechanical medium and the vacuum medium*

| | **Mechanical medium** *(Elastic solid)* | **Vacuum medium** | |
|---|---|---|---|
| | | *Electric component* | *Magnetic component* |
| **Strain** | **r** | Electric displacement **D** | Magnetic flux **B** |
| **Stress** | $f \propto \mathbf{r}$<br>$\sigma_{ij} = \lambda r_{k,k}\delta_{ij} + \mu(r_{i,j} + r_{j,i})$ | Electric field<br>$\mathbf{E} = \frac{1}{\varepsilon_0}\mathbf{D}$ | Magnetic field<br>$\mathbf{H} = \frac{1}{\mu_0}\mathbf{B}$ |
| **Helmholtz decomposition** | $\mathbf{r} = -\nabla\phi + \nabla\times\boldsymbol{\psi}$ | $\mathbf{D} = \nabla\times\mathbf{Z}$ | $\mathbf{B} = \nabla\times\mathbf{A}$ |
| **Potential function** | Dilational wave $\phi$ and transverse wave $\psi$ | Electric vector potential **Z** | Magnetic vector potential **A** |
| **Cross-interacting mechanism:**<br><br>**Coupling equation** | Newton's law<br><br>Generalized Hooke's law | $\frac{\partial \mathbf{E}}{\partial t} = \frac{1}{\varepsilon_0}\nabla\times\mathbf{H}$<br>Ampere's law<br>(as modified by Maxwell)<br>$\frac{\partial \mathbf{Z}}{\partial t} = \frac{1}{\mu_0}\nabla\times\mathbf{A}$ (42) | $\frac{\partial \mathbf{H}}{\partial t} = -\frac{1}{\mu_0}\nabla\times\mathbf{E}$<br>Faraday's law<br><br>$\frac{\partial \mathbf{A}}{\partial t} = -\frac{1}{\varepsilon_0}\nabla\times\mathbf{Z}$ (40) |
| **Wave equation (transverse)** | $\nabla^2\boldsymbol{\psi} - \frac{1}{c_s^2}\frac{\partial^2\boldsymbol{\psi}}{\partial t^2} = 0$<br>where $c_s = \sqrt{\mu/\rho}$ | $\nabla^2\mathbf{Z} - \frac{1}{c^2}\frac{\partial^2\mathbf{Z}}{\partial t^2} = 0$<br>where $c = 1/\sqrt{\mu_0\varepsilon_0}$ | $\nabla^2\mathbf{A} - \frac{1}{c^2}\frac{\partial^2\mathbf{A}}{\partial t^2} = 0$<br>where $c = 1/\sqrt{\mu_0\varepsilon_0}$ |
| **Wave equation (longitudinal)** | $\nabla^2\phi - \frac{1}{c_p^2}\frac{\partial^2\phi}{\partial t^2} = 0$<br>where $c_p = \sqrt{(\lambda+2\mu)/\rho}$ | Not applicable | Not applicable |

From Table 2, it is also clear that no matter for a mechanical medium or for the vacuum medium, wave propagation requires a cross-interacting mechanism. For wave propagation in an elastic solid, the cross-interaction is mediated through two coupling equations, i.e., the Newton's law and the generalized Hooke's law. For the vacuum medium, the coupling equations appear to be the Ampere's law (as modified by Maxwell) and Faraday's law. But at a deeper level, we find the coupling equations are more like the cross-interactions between the electric vector potential **Z** and the magnetic vector potential **A**, i.e., Eqs. (40) and (42). It is their cross-interactions that generated the excitation wave.

Another interesting point one can see from Table 2 is that, no matter for mechanical medium or vacuum medium, all the wave equations appear to have the same form. They are highly symmetrical. For later reference, we may call these wave equations "*4-dimensional Laplace equation*". We may point out that, although the wave equations in different systems look the same, their wave functions represent very different *basic fields*. Such fields are associated with different measurements of the medium displacement.

# 6. Derivation of the quantum mechanical equations based on the matter wave model: Derivation of the Klein-Gordon equation

Since we propose that all free particles are excitation waves of the vacuum, and different particles are associated with different excitation modes, one may expect that the wave equation for all types of particle should be derivable from the basic wave equation of the matter wave. The question now is that: How can this model reconcile with the conventional theories in quantum mechanics? In the conventional quantum theory, different types of particles are described by different wave equations. For example, the motion of a scalar particle (spin = 0) is described by the Klein-Gordon equation; the motion of an electron (spin = ½ ), is described by the Dirac equation under the relativistic condition, or by the Schrödinger equation under the non-relativistic condition. Are these quantum wave equations related to wave equation of the matter wave?

In the following sections, we will show that all conventional quantum wave equations can indeed be derived from the basic wave equation of matter wave. Let us consider first the movement of a scalar particle with rest mass. This is traditionally described by the Klein-Gordon equation in relativistic quantum mechanics. In the following, we will show that the Klein-Gordon equation can be derived directly based on the matter wave model.

### 6.1. What is the basic wave equation of a free particle?

As we stated at the beginning of this paper, we believe that both matter wave and radiation wave are excitation waves of the same vacuum medium. Then, their wave equations should be the same. From the discussions in the last section, we know the vacuum is like a dielectric medium. The basic field of the excitation wave can be identified as the electric vector potential **Z** which is associated with the electric displacement of the vacuum. Its wave equation is the 4-dimensional Laplace equation,

$$\nabla^2 \mathbf{Z} - \frac{1}{c^2} \frac{\partial^2 \mathbf{Z}}{\partial t^2} = 0. \tag{43A}$$

Then, we may assume that the wave function of the matter wave ($\psi$) is directly equal to $\mathbf{Z}$, or $\psi$ is a linear function of $\mathbf{Z}$. In either case, $\psi$ should satisfy the same wave equation as $\mathbf{Z}$, that is,

$$\nabla^2 \psi - \frac{1}{c^2} \frac{\partial^2 \psi}{\partial t^2} = 0. \tag{44}$$

We will regard this as the *basic wave equation* (BWE) of matter wave.

One may question our assumption by pointing out that there is a fundamental difference between radiation wave and matter wave. That is, radiation wave describes only the movement of electromagnetic wave (photon) which has no rest mass. Matter wave, on the other hand, represents a massive particle (such as an electron) which apparently has rest mass ($m$). Thus, one may expect that the wave equation for a matter wave should contain the parameter $m$. In fact, this is indeed the practice used in the conventional quantum theory; the wave equations for describing the motion of massive particles, such as the Klein-Gordon equation or the Dirac equation, all contain explicitly the parameter $m$.

We think this argument is not totally correct. We believe **mass should be treated on the same footing as energy and momentum**. According to our hypothesis, all free particles are excitation waves of the *same* vacuum medium; thus, their basic wave equation is determined only by the physical property of the transmission medium. Since different solutions of the basic wave equation represent different types of particles, $m$ should not appear in the basic wave equation; it should emerge only when a particular solution is chosen. In another word, $m$ should appear as an eigenvalue for the solution of the wave equation. This is similar to the case that the energy level $E$ does not appear in the time-dependent Schrödinger equation. $E$ only appears in the time-independent Schrödinger equation [17].

It also means that, none of the equations containing $m$ as an explicit parameter (including the Klein-Gordon equation, the Schrödinger equation and the Dirac equation) can be regarded as the basic wave equation for matter wave.

### 6.2. Solutions of the wave equation

The simplest solution of the basic wave equation, Eq. (44), is a plane wave

$$\psi_{\hat{k}} \propto e^{i(\mathbf{k} \cdot \mathbf{x} - \omega t)}, \tag{45}$$

where $\mathbf{k}$ and $\omega$ are the wave vector and frequency, respectively. This plane wave solution is commonly used to represent the wave function of a photon moving along the direction $\hat{k}$. Can this plain wave solution also be used to represent the wave function for a free particle with rest mass? We think the answer is "no". Since a particle with rest mass would behave like a pointed object macroscopically, the probability of detecting the particle (i.e., $|\psi|^2$) is expected to be highest at its trajectory. Thus, its wave function should depend not only on the coordinate parallel to its trajectory (i.e., $\hat{k} \cdot x$), but also on the coordinates in the transverse plane ($\hat{k} \times x$). That means the matter wave representing a free particle must have a cylindrical symmetry, i.e.,

$$\psi_{\hat{k}}(x,t) = \psi_T(\hat{k} \times x) \psi_{path}(\hat{k} \cdot x, t), \tag{46}$$

where $\psi_L$ is the longitudinal component of the wave function which describes the travelling wave along the particle's trajectory, and $\psi_T$ is the transverse component of the wave function in the transverse plane. Substituting Eq. (46) into Eq. (44), and using the technique of separation of variables, Eq. (44) can be converted into two simultaneous equations [36],

$$\begin{cases} \left[\nabla^2 - \frac{1}{c^2}\frac{\partial^2}{\partial t^2}\right]\psi_{path}(\hat{k} \cdot x, t) = \ell^2 \psi_{path}(\hat{k} \cdot x, t) & (47) \\ \nabla^2 \psi_T(\hat{k} \times x) = -\ell^2 \psi_T(\hat{k} \times x), & (48) \end{cases}$$

where $\ell^2$ is a coupling constant. The above equations can be solved separately. The solution of Eq. (47) is a plane wave

$$\psi_{path}(\hat{k} \cdot x, t) \propto e^{i(k \cdot x - \omega t)}, \tag{49}$$

where $k = k\hat{k}$ and

$$\omega = (k^2 + \ell^2)^{1/2} c. \tag{50}$$

The solution of Eq. (48) is

$$\psi_T(\hat{k} \times x) \propto J_n(\ell r) e^{\pm in\theta}, \tag{51}$$

where $J_n$ is Bessel function of the first kind (with *n* as an integer or a half integer); *r* and $\theta$ are the radius and the azimuthal angle in the transverse plane. To simplify our notation, let us define the direction of particle movement as the *z*-axis, i.e., $\hat{k} \parallel \hat{z}$. Then, $k \cdot x = kz$. Substituting Eqs. (49), (51) into Eq. (46), the wave function of a free particle is equal to

$$\psi_{\hat{k}}(x, t) = a J_n(\ell r) e^{\pm in\theta} e^{i(kz - \omega t)}, \tag{52}$$

(where *a* is a normalizing constant). As expected, the wave function of a free particle behaves like a travelling wave along the direction of its trajectory. Because of the phase factor $e^{\pm in\theta}$ and the Bessel function $J_n(\ell r)$, $\psi_{\hat{k}}$ propagates in a helical fashion and decreases in an oscillating manner in the transverse direction. It behaves almost like a vortex. (See Figure 3).

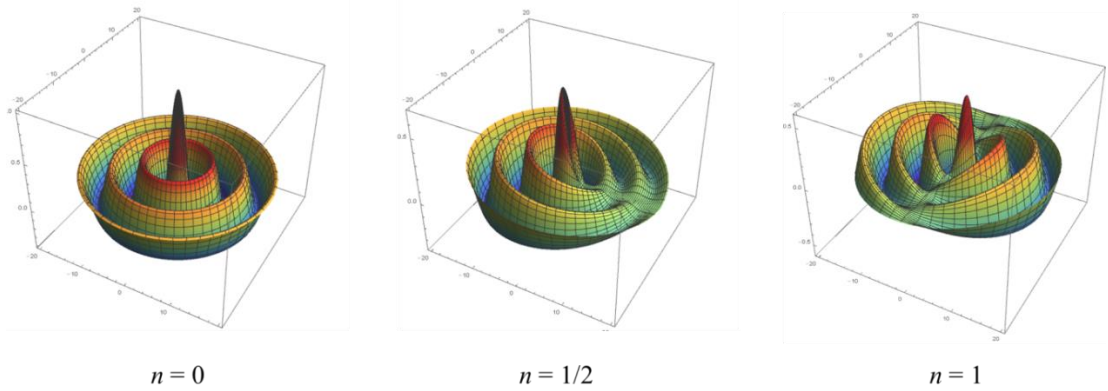

| n = 0 | n = 1/2 | n = 1 |

*Figure 3. A 3-D plot of the wave function $\psi_{\hat{k}}$ for n = 0, ½, 1. The direction of wave propagation is along the z-axis.*

### 6.3. Identifying the physical meaning of parameters within the wave function: Correspondence between wave properties and particle properties

The wave function shown in Eq. (52) contains four parameters, $\omega$, **k**, $\ell$ and $n$. What are their physical meanings? We can get some idea by comparing this wave function with that of a photon. In the case of a photon wave function (Eq. (45)), it is well known that $\omega$ and **k** are related to the energy ($E$) and momentum (**p**) of the particle, as described by the Planck's relation and the de Broglie relation. If we compare the traveling wave component of Eq. (52) with Eq. (45), it is easy to see that $\omega$ and **k** should have the same meanings, i.e., $E = \hbar\omega$ and $\mathbf{p} = \hbar\mathbf{k}$. (In fact, if one applies the correspondence rules by using Eq. (52), one can show that both the Planck's relation and de Broglie's relation hold for a wave function representing a free particle with mass) [32].

Now, what is the physical meaning of $\ell$? Our recent work suggested that $\ell$ is related to the rest mass of the particle [31]. This can be easily shown. From the Planck's relation and de Broglie's relation, Eq. (50) becomes

$$E^2 = c^2\left(p^2 + \hbar^2\ell^2\right). \tag{53}$$

Recall that the particle velocity ($v$) is determined by the group velocity of the wave packet [6], $v = \dfrac{\partial \omega}{\partial k} = \dfrac{\partial E}{\partial p}$, and the particle mass $M$ is defined by $p = Mv$ in the classical limit, one can show from Eq. (53) that

$$M = \frac{\hbar\ell/c}{\left(1 - v^2/c^2\right)^{1/2}}. \tag{54}$$

At $v \ll c$, $M$ equals the rest mass, $m$. That means the rest mass is

$$m = \frac{\hbar \ell}{c} . \tag{55}$$

This result makes good sense, since when we substitutes Eq. (55) into Eq. (54), we have

$$M = \frac{m}{\left(1 - v^2/c^2\right)^{1/2}} . \tag{56}$$

This is a known result from earlier experimental observations, which showed that the particle mass is speed-dependent [37]. Furthermore, by substituting Eq. (55) into Eq. (53), we have

$$E^2 = p^2 c^2 + m^2 c^4 , \tag{57}$$

which is identical to the relativistic relations between energy and momentum [38]. Furthermore, by combining Eqs. (56), (57), and recall that $p = Mv$, one can obtain

$$E = Mc^2 . \tag{58}$$

At $v \ll c$, the above equation becomes $E = mc^2$. Thus, by identifying the wave parameter $\ell$ with the rest mass $m$, one can naturally obtain the known relationship between energy and mass.

Our finding that the wave parameter $\ell$ is associated with the rest mass is not totally surprising. Since both momentum and energy of a free particle are known to be connected with "wave vector/wave number" (inverse of wavelength) in the spatial and temporal dimensions, it is reasonable to speculate that the rest mass may be connected with some sort of "intrinsic wave number" too. This is indeed the case. The asymptotic form of the Bessel function is known to be

$$J_n(\ell r) \to \left(\frac{2}{\pi \ell r}\right)^{1/2} \cos\left(\ell r - \frac{2n+1}{4}\pi\right) . \tag{59}$$

Thus, $\ell$ can be regarded as the "transverse wave number" of the free particle. In fact, from Eqs. (55) and (59), one can easily see that the wavelength of this transverse oscillation is

$$\lambda = \frac{2\pi}{\ell} = \frac{h}{mc} , \tag{60}$$

which is identical to the so-called "Compton wavelength" ($\lambda_c$) of the particle [39].

Our finding that the rest mass is associated with the oscillation periodicity in the transverse direction appears to make very good sense. It is closely parallel to the Planck's relation and the de Broglie relation, which show that the energy and momentum are related to the periodicity of oscillation of the vacuum medium. More specifically, $E$ is shown to be related to the periodicity of oscillation in the time dimension, while $p$ is related to the periodicity of oscillation in the spatial dimension along the direction of the particle trajectory. In essence, these results suggest that, *energy, momentum and mass have very similar physical meanings; all of them are measures of the curvature of bending the vacuum medium along different dimensions*.

## 6.4. Derivation of the Klein-Gordon equation from the BWE

Now we can see the Klein-Gordon equation is directly related to the BWE. In Section 6.2, we have pointed out that the matter wave can be decoupled into a longitudinal component and a transverse component,

$$\psi_{\hat{k}}(x,t) = \psi_T(\hat{k} \times x)\, \psi_{path}(\hat{k} \cdot x,\, t). \tag{46}$$

Using the technique of "separation of variables", one can show that the BWE for the matter wave can be separated into two coupled equations, i.e., Eqs. (47) and (48). The equation of motion for the longitudinal component is

$$\left[\nabla^2 - \frac{1}{c^2}\frac{\partial^2}{\partial t^2}\right] \psi_{path}(\hat{k} \cdot x,\, t) = \ell^2 \psi_{path}(\hat{k} \cdot x,\, t). \tag{47}$$

As shown in the last Section, the wave parameter $\ell$ is found to be connected with the rest mass $m$, such that $m = \hbar \ell / c$. Then, we can re-write Eq. (47) as

$$(\nabla^2 - \frac{1}{c^2}\frac{\partial^2}{\partial t^2})\psi_{path} - \left(\frac{mc}{\hbar}\right)^2 \psi_{path} = 0\ . \tag{61}$$

This equation is identical to the "Klein-Gordon equation" [40]

$$(\nabla^2 - \frac{1}{c^2}\frac{\partial^2}{\partial t^2})\phi - \left(\frac{mc}{\hbar}\right)^2 \phi = 0\ , \tag{62}$$

if we identify its wave function $\phi$ with $\psi_{path}$. This means that, *the Klein-Gordon equation can indeed be derived from the BWE*. From this derivation, it is also clear that *the wave function of the Klein-Gordon equation describes only the longitudinal component of the matter wave*, i.e., the motion of the particle along its trajectory ($\psi_{path}$).

## 7. Derivation of the Dirac equation from the BWE

As we mentioned earlier, since all free particles are excitation waves of the vacuum medium, they should obey the same wave equation regardless of mass and spin. This wave equation should depend only on the physical properties of the vacuum medium. In Sections 5 and 6, we have identified that this common wave equation is the BWE; different particles are represented by its different solutions. Since the BWE is a general wave equation for all particles, it is not only valid for scalar particles (which satisfy the Klein-Gordon equation), the BWE should also be valid for spin = ½ particles.

This means that one should be able to derive the wave equations for an electron from BWE. In 1928, Dirac showed that an electron can be described by a relativistic quantum mechanical equation, which is now called the "Dirac equation" [23], [41]. In the following, we will show that the BWE can indeed lead to the Dirac equation.

### 7.1. Deriving the Dirac equation by linearizing the Klein-Gordon equation

In order to derive the Dirac equation from the BWE, we make use of three considerations:

(a) We hypothesis that, like the Klein-Gordon equation, the Dirac equation may represent only the longitudinal component of the matter wave. That is, the wave function of the Dirac equation is connected to the $\psi_{path}$ component of $\psi_{\hat{k}}$.

(b) If this is the case, one may be able to show that the Dirac equation is a special case of the Klein-Gordon equation. Since we already knew the Klein-Gordon equation is a special case of the BWE, then, the solution of the Dirac equation should automatically satisfy the BWE.

(c) The wave function representing a matter wave is not necessary to be a scalar function. Depending on the spin of the particle, the wave functions may have different mathematical forms. Some can be scalars, while others can be vectors or matrices. For particles with spin = 0, the solution can simply be a scalar function. In the case of an electron, the situation could be more complicated. Since it has a spin = ½, its full wave function (as shown in Eq. (52)) cannot be a single-valued function in respect to $\theta$. Thus, it may be more appropriate to use a matrix to represent it.

If we denote $\psi$ as a special solution of the Klein-Gordon equation that represents a free electron, $\psi$ should satisfy

$$(\nabla^2 - \frac{1}{c^2}\frac{\partial^2}{\partial t^2})\psi - \left(\frac{mc}{\hbar}\right)^2 \psi = 0 \ . \tag{63}$$

To simplify the mathematical calculation, let us use the natural unit (i.e., $\hbar = 1$, $c = 1$) for the above equation,

$$-\frac{\partial^2 \psi}{\partial t^2} + \nabla^2 \psi - m^2 \psi = 0. \tag{64}$$

The left-hand side of this equation can be decomposed into the product of two factors

$$\left(i\frac{\partial}{\partial t} - i\alpha \cdot \nabla + \beta m\right)\left(i\frac{\partial}{\partial t} + i\alpha \cdot \nabla - \beta m\right)\psi = 0. \tag{65}$$

Then, the above equation can be decoupled into two independent equations,

$$\begin{cases} \left(i\frac{\partial}{\partial t} + i\alpha \cdot \nabla - \beta m\right)\psi = 0 & (66) \\ \left(i\frac{\partial}{\partial t} - i\alpha \cdot \nabla + \beta m\right)\psi = 0. & (67) \end{cases}$$

Eq. (65) can be rewritten explicitly as

$$\left[-\frac{\partial^2}{\partial t^2}+(\alpha\cdot\nabla)^2+im\beta(\alpha\cdot\nabla)+im(\alpha\cdot\nabla)\beta-\beta^2 m^2\right]\psi=0. \tag{65A}$$

One can see that, in order to make Eq. (65A) equal to Eq. (64), the parameter $\alpha$ and $\beta$ must satisfy the following conditions:

$$\begin{cases} \beta^2=1 \\ \alpha\beta+\beta\alpha=0 \\ \alpha_i\alpha_j=1,\text{ if } i=j \\ \alpha_i\alpha_j+\alpha_j\alpha_i=0,\text{ if } i\neq j. \end{cases} \tag{68}$$

These conditions post a restriction on the possible mathematical form of the parameters $\alpha$ and $\beta$. If $\alpha$ and $\beta$ are ordinary numbers, it is impossible for them to satisfy Eqs. (68). In order to satisfy those conditions, both $\alpha$ and $\beta$ must be treated as a 4 x 4 matrix. It can be shown that the conditions Eq. (68) can indeed be satisfied if one defines the matrices $\alpha$ and $\beta$ as [23]

$$\alpha_i=\begin{pmatrix}-\sigma^i & \mathbf{0}\\ \mathbf{0} & \sigma^i\end{pmatrix},\ \beta=\begin{pmatrix}\mathbf{0} & \sigma^o\\ \sigma^o & \mathbf{0}\end{pmatrix},$$

where the various $\sigma^i$ are $2\times 2$ Pauli spin matrices,

$$\sigma^o=\begin{pmatrix}1 & 0\\ 0 & 1\end{pmatrix},\ \sigma^1=\begin{pmatrix}0 & 1\\ 1 & 0\end{pmatrix},\ \sigma^2=\begin{pmatrix}0 & -i\\ i & 0\end{pmatrix},\ \sigma^3=\begin{pmatrix}1 & 0\\ 0 & -1\end{pmatrix},$$

and
$$\mathbf{0}=\begin{pmatrix}0 & 0\\ 0 & 0\end{pmatrix}.$$

The linearized equation (Eq. (66)) is called the "Dirac equation". From the above analysis, it is clear that, if one wants to make the Dirac equation to satisfy the Klein-Gordon equation, $\alpha$ and $\beta$ must be treated as 4 x 4 matrices. In another word, the Dirac equation must be treated as 4 x 4 matrix equation.

### 7.2. Meaning of the Dirac wave function

It is now clear that the Dirac equation is a special case of the Klein-Gordon equation, which in turn is a special case of the BWE. Thus, any solution of the Dirac equation should also satisfy the BWE. This means that all electron wave functions obtained from the Dirac equation should automatically satisfy the basic wave equation of the vacuum.

Since the Dirac equation is a matrix equation, the wave function $\psi$ must become a four-component column matrix (which is called a "spinor field").

$$\psi = \begin{pmatrix} \psi_L \\ \psi_R \end{pmatrix} = \begin{pmatrix} \psi_1 \\ \psi_2 \\ \psi_3 \\ \psi_4 \end{pmatrix} \qquad (69)$$

where

$$\psi_L = \begin{pmatrix} \psi_1 \\ \psi_2 \end{pmatrix}, \; \psi_R = \begin{pmatrix} \psi_3 \\ \psi_4 \end{pmatrix}. \qquad (70)$$

In current literature, the four component column matrix $\psi$ is called the "*Dirac spinor*"[41], $\psi_L$ is called the *left-hand spinor* and $\psi_R$ is called the *right-hand spinor*. The Dirac equation has been shown to be highly successful in describing the motion of an electron. In fact, the modern quantum field theory of electrons is almost entirely based on the Dirac equation. From the above review, one can see clearly that the Dirac equation is a derivative of the Klein-Gordon equation. But since the Klein-Gordon equation itself is a derivative of the BWE, it is obvious that the Dirac equation is also a derivative of the BWE. Thus, the wave function representing an electron is just a special solution of the BWE of the matter wave.

Why the wave function representing an electron is a spinor instead of a scalar? It is probably because it is a particle with $n = 1/2$. This makes the wave function not returning to the original values when $\theta$ is rotated for $2\pi$. That means the wave function cannot be a single-valued function. It should have at least two values: (a) $\psi(\theta = 0 \text{ to } 2\pi)$; (b) $\psi(\theta = 2\pi \text{ to } 4\pi)$. This implies that $\psi$ can be a column matrix with at least two components. But since the phase for the wave equation can take on $\pm in\theta$, the wave function can rotate either clockwise or counter-clockwise. The wave function could be a mixture of a left-hand matrix and a right-hand matrix. This may explain why the wave function representing a free electron should have four components as shown in Eq. (69).

## 8. Derivation of the Schrödinger equation from the BWE

In the historical development of quantum mechanics, the first wave equation proposed for electron was not the Dirac equation; it was the Schrödinger equation. In fact, the Schrödinger equation is far more widely used today. One may say that the foundation of atomic physics, molecular physics and solid state physics are all based on the Schrödinger equation.

Although the appearance of the Dirac equation and the Schrödinger equation are different, they are actually closely connected. It is known that the Schrödinger equation can be reduced from the Dirac equation when the potential and kinetic energies of the electron is much smaller than its rest mass [42]. Thus, the wave functions from these two equations are closely related.

In this work, we will not attempt to derive the Schrödinger equation from the Dirac equation. Instead, we will derive the Schrödinger equation based on the Klein-Gordon equation. Such an approach is not only more faithful to history, one can also see more clearly the physical meaning of the wave function in the Schrödinger equation.

## 8.1. Physical basis of the correspondence rules

In order to derive the Schrödinger equation, we need to first introduce the "*correspondence rules*", which played an important role in Schrödinger's original derivation. From the matter wave model proposed here, one can easily see the rationale behind these rules. To make things simple, let us first use the radiation wave (photon) as an example. As we showed earlier, a simple solution for the BWE is the wave function of a photon, which is assumed to be a plane wave,

$$\psi_{\hat{k}} \propto e^{i(\mathbf{k} \cdot \mathbf{x} - \omega t)}. \tag{45}$$

Based on the Planck's relation, we know

$$E = \hbar \omega. \tag{71}$$

Using the above two equations, one can see that

$$\left(i\hbar \frac{\partial}{\partial t}\right)\psi_{\hat{k}} = E\psi_{\hat{k}}. \tag{72}$$

This relation suggests that, if one wants to associate the particle property $E$ with an operator on the wave function, the proper correspondence rule should be

$$E \rightarrow i\hbar \frac{\partial}{\partial t}. \tag{73}$$

Similarly, recall from the de Broglie relation, $\mathbf{p} = \hbar \mathbf{k}$. Using Eq. (45), one can see that

$$\left(\frac{\hbar}{i}\nabla\right)\psi_{\hat{k}} = \mathbf{p}\psi_{\hat{k}}. \tag{74}$$

This relation suggests that, if one wants to associate the momentum $\mathbf{p}$ with an operator on the wave function, the proper correspondence rule should be

$$\mathbf{p} \rightarrow \frac{\hbar}{i}\nabla. \tag{75}$$

These correspondence rules are summarized in Table 3. It can be demonstrated that these correspondence rules make very good sense. For the photon, we know it obeys the mechanical relation $E^2 = c^2 p^2$ in a particle view. Using the above correspondence rules, one can write down its wave equation as

$$\left(i\hbar \frac{\partial}{\partial t}\right)^2 \psi_{\hat{k}} = c^2 \left(\frac{\hbar}{i}\nabla\right)^2 \psi_{\hat{k}}.$$

This means that

$$\nabla^2 \psi_{\hat{k}} - \frac{1}{c^2} \frac{\partial^2 \psi_{\hat{k}}}{\partial t^2} = 0, \qquad (76)$$

which is exactly the well-known wave equation of a photon. This is a simple demonstration that by using the correspondence rules, one can easily determine the wave equation of a particle based on its mechanical relation.

*Table 3. Correspondence rules for both radiation wave and matter wave*

| Wave function | Physical Law | Operation on the wave function | Correspondence rule | Mechanical relation |
|---|---|---|---|---|
| **Radiation wave (photon)** $\psi_{\hat{k}} \propto e^{i(\mathbf{k}\cdot\mathbf{x}-\omega t)}$ | Planck relation $E = \hbar\omega$ | $E\psi_{\hat{k}} = \left(i\hbar\dfrac{\partial}{\partial t}\right)\psi_{\hat{k}}$ | $E \to i\hbar\dfrac{\partial}{\partial t}$ | $E^2 = c^2 p^2$ |
| | de Broglie relation $\mathbf{p} = \hbar\mathbf{k}$ | $\mathbf{p}\psi_{\hat{k}} = \left(\dfrac{\hbar}{i}\nabla\right)\psi_{\hat{k}}$ | $\mathbf{p} \to \dfrac{\hbar}{i}\nabla$ | |
| **Matter wave (path component)** $\psi_{path}(\hat{\mathbf{k}}\cdot\mathbf{x},t) \propto e^{i(\mathbf{k}\cdot\mathbf{x}-\omega t)}$ | Planck relation $E = \hbar\omega$ | $E\psi_{path} = \left(i\hbar\dfrac{\partial}{\partial t}\right)\psi_{path}$ | $E \to i\hbar\dfrac{\partial}{\partial t}$ | $E^2 = p^2c^2 + m^2c^4$ |
| | de Broglie relation $\mathbf{p} = \hbar\mathbf{k}$ | $\mathbf{p}\psi_{path} = \left(\dfrac{\hbar}{i}\nabla\right)\psi_{path}$ | $\mathbf{p} \to \dfrac{\hbar}{i}\nabla$ | |

The above correspondence rules not only hold for radiation waves (photon), they can also hold for matter wave representing a free particle. If one considers only the path component of the matter wave ($\psi_{path}$), one will find the argument used for photon can also be applied for the matter wave. (See Table 3). Again, one can show that, by using the correspondence rules, one can easily identify the wave equation of the matter wave based on its mechanical relations. In Section 6.3, we have demonstrated that the energy-momentum relationship for a free particle is

$$E^2 = p^2 c^2 + m^2 c^4 . \qquad (57)$$

Applying the correspondence rules shown in Table 3, the corresponding wave equation for the above will become

$$\left(i\hbar\frac{\partial}{\partial t}\right)^2 \phi = c^2 \left(\frac{\hbar}{i}\nabla\right)^2 \phi + m^2 c^4 \phi .$$

By simple re-arrangement, one can see that the above equation is exactly the Klein-Gordon equation,

$$(\nabla^2 - \frac{1}{c^2}\frac{\partial^2}{\partial t^2})\phi - \left(\frac{mc}{\hbar}\right)^2 \phi = 0 \ . \tag{62}$$

Thus, it is clear that the correspondence rules as shown in Table 3 can be applied to both radiation waves and matter waves.

### 8.2. Derivation of the Schrödinger equation based on the Klein-Gordon equation

With the establishment of the correspondence rules, one can show that the Schrödinger equation can be derived from the Klein-Gordon equation. As shown in the last sub-Section, we know the Klein-Gordon equation

$$(\nabla^2 - \frac{1}{c^2}\frac{\partial^2}{\partial t^2})\phi - \left(\frac{mc}{\hbar}\right)^2 \phi = 0 \tag{62}$$

is corresponding to the mechanical relation between energy and momentum,

$$E^2 = c^2\left(p^2 + m^2 c^2\right). \tag{57}$$

At the low-speed limit, $v \ll c$, $p = Mv \ll mc$, the above relation becomes

$$E = mc^2\left(1 + \frac{p^2}{m^2 c^2}\right)^{1/2} = mc^2\left(1 + \frac{1}{2}\frac{p^2}{m^2 c^2} + ...\right) \approx mc^2 + \frac{p^2}{2m}. \tag{77}$$

Using the correspondence rules, we can see that the non-relativistic wave equation should be

$$i\hbar \frac{\partial \phi}{\partial t} = \left(mc^2 - \frac{1}{2m}(-i\hbar\nabla)^2\right)\phi. \tag{78}$$

This is the non-relativistic counter-part of the Klein-Gordon equation. It is the low-speed wave equation for a free particle.

In order to apply this wave equation to describe the movement of an electron in an external electric field, we need to make a small modification of Eq. (78). The electron has an electric charge $q$. Suppose the electron is inside an atom, where there is a negative electrical potential $V = V(r)$, its energy is shifted by an amount $-qV$, i.e.,

$$E \rightarrow E - qV \rightarrow i\hbar \frac{\partial}{\partial t} - qV \ .$$

Eq. (78) will then become

$$\left(i\hbar \frac{\partial}{\partial t} - qV\right)\phi = \left(mc^2 - \frac{\hbar^2 \nabla^2}{2m}\right)\phi. \tag{79}$$

Now, let us define a new wave function by applying a global transformation,

$$\psi_s = e^{imc^2 t/\hbar}\phi. \tag{80}$$

By substituting Eq. (80) into Eq. (79), the $mc^2$ term can be canceled out. Then, we have

$$i\hbar \frac{\partial \psi_s}{\partial t} = \left(-\frac{\hbar^2}{2m}\nabla^2 + qV\right)\psi_s. \tag{81}$$

This is exactly the Schrödinger equation for an electron in the presence of an electrical potential.

### 8.3. Meaning of the wave function in the Schrödinger equation

From the above derivation, one can immediately see that:

(a) The wave function of the Schrödinger equation ($\psi_s$) is directly related to the wave function of the Klein-Gordon equation ($\phi$); they differ only by a phase factor as shown in Eq. (80).

(b) Because $\psi_s$ is connected to $\phi$, and we know that $\phi$ represents the path component of the matter wave, i.e., $\phi = \psi_{path}$, $\psi_s$ is only associated with the longitudinal component of the matter wave which describes the motion of the free particle along its pathway ($\psi_{path}$).

(c) Since $\psi_s$ differs from $\phi$ by a phase factor $e^{imc^2 t/\hbar}$, $\psi_s$ does not oscillate as rapidly as $\phi$. Instead, $\psi_s$ oscillates at a lower frequency (at $\omega = E/\hbar$).

In summary, based on the above discussions, we see that all known quantum mechanical wave equations are derivatives of the BWE. In another word, particles of different spins (including photons, scalar particles and electrons) are all excitation waves of the same vacuum medium and they all obey the BWE.

## 9. Discussions

### 9.1. There is a natural transition from classical mechanics to quantum mechanics

In many textbooks, quantum mechanics and classical mechanics are often treated on separated grounds; they are supposed to be based on very different physical principles. In fact, the physical basis of quantum phenomena is often regarded as a mystery. Some people even said that perhaps it is beyond human capability to explain the magic of quantum mechanics [43], [44]. We do not agree with this pessimistic view. In this work, we try to show that it is possible to understand the principle behind quantum mechanics by examining its transition from classical mechanics. We propose a model that treats all particles as excitation waves of the vacuum. The quantum wave equation then can be derived based on the physical properties of the vacuum medium, just like the mechanical wave equations can be derived from the physical properties of the mechanical medium using well established physical laws.

*Table 4. Comparison between wave mechanisms in the mechanical medium and in the vacuum medium*

| Wave type | Medium | Physical Law | Wave equation | Wave function |
|---|---|---|---|---|
| **Mechanical wave (sound)** | Mechanical medium | Newton's law $F = ma = m\dfrac{d^2 x}{dt^2}$ | Harmonic oscillator: $m\dfrac{d^2 x}{dt^2} = -\kappa x$ | $x = x_0 e^{i\omega t}$ where $\omega = \sqrt{\kappa/m}$ |
| | | | 1-D string $\dfrac{\partial^2 \phi}{\partial x^2} - \dfrac{1}{c_1^2}\dfrac{\partial^2 \phi}{\partial t^2} = 0$ | $\phi = \phi_0 e^{i(z - c_1 t)}$ where $c_1 = \sqrt{F_1/\rho}$ |
| | | Hooke's law $F = -\kappa x$ | Elastic solid: $\nabla^2 \psi - \dfrac{1}{c_s^2}\dfrac{\partial^2 \psi}{\partial t^2} = 0$ | $\psi \propto e^{i(\mathbf{k}\cdot\mathbf{x} - \omega t)}$ where $c_s = \omega/k$ |
| **Radiation wave (photon)** | Vacuum | Maxwell's equation | $\nabla^2 \psi_{\hat{k}} - \dfrac{1}{c^2}\dfrac{\partial^2 \psi_{\hat{k}}}{\partial t^2} = 0$ where $c = 1/\sqrt{\mu_0 \varepsilon_0}$ | $\psi_{\hat{k}} \propto e^{i(\mathbf{k}\cdot\mathbf{x} - \omega t)}$ |
| **Matter wave (massive particles)** | Vacuum | Cross-interaction between **A** and **Z** | $\nabla^2 \psi - \dfrac{1}{c^2}\dfrac{\partial^2 \psi}{\partial t^2} = 0$ where $c = 1/\sqrt{\mu_0 \varepsilon_0}$ | $\psi(\mathbf{x}, t) = a J_n(\ell r) e^{\pm i n \theta} e^{i(kz - \omega t)}$ |

In the case of the mechanical systems, the wave mechanics is described by the Newton's laws and generalized Hooke's law. In the vacuum medium, the wave mechanism is described by the Maxwell's equations (Ampere's law and Faraday's law). More specifically, we showed that the excitation wave of the vacuum is driven by a cross-interaction between the electric vector potential **Z** and the magnetic vector potential **A**. By studying the transition from classical mechanics to quantum mechanics, one can obtain a clear understanding on the physical meaning of the wave function. In the classical mechanical system (i.e., elastic solid), the wave function apparently represents a potential function which is associated with the displacement of the transmitting medium according to the Helmholtz decomposition theorem. Similarly, the wave function of a particle is also related to a potential function, which is associated with the displacement of the vacuum medium based on the Helmholtz decomposition. Thus, there is a close analogy between the wave mechanisms in a classical system and that in a quantum mechanical system. (See Table 4).

**9.2. All quantum wave equations can be traced to the BWE**

In this work, we showed that in both the mechanical medium and the vacuum medium, the wave equations always appear as a 4-D Laplace equation. We call this equation the "*Basic Wave Equation*" (BWE). We think all free particles are excitation waves of the vacuum medium, and different particles are just different excitation modes. If the matter wave is described by the BWE, then different solutions of the BWE would represent different particles. These particles could have different mass or spin. But their wave functions should all satisfy the BWE. In the conventional theories, different particles are thought to satisfy different quantum wave equation. For example, the photon is supposed to satisfy the wave equation of light, the scalar particle is supposed to satisfy the Klein-Gordon equation, and the electrons are supposed to satisfy the Dirac equation (or the Schrödinger equation). In order to reconcile our matter wave model with the conventional quantum theories, one must demonstrate that all conventional quantum equations for different particles are derivable from the BWE. In Sections 7 to 9 of this paper, we showed this is indeed the case. In another word, one could say that *the BWE is the mother of all quantum wave equations*. To demonstrate this point more clearly, we have summarized the relationship between the wave equations and wave functions of different types of free particles in Table 5.

*Table 5. Physical properties of the wave function for different free particles*

| Particle type | Rest mass | Spin | Mathematical form of $\psi$ | Nature of $\psi$ | Wave equation |
|---|---|---|---|---|---|
| Photon | No | 1 | Vector | A special solution of BWE | Wave equation of light (by Maxwell) |
| Scalar particle | Yes | 0 | Scalar | A special solution of BWE | Klein-Gordon equation |
| Lepton/Electron (relativistic) | Yes | ½ | Spinor | A special solution of BWE | Dirac equation |
| Electron (non-relativistic) | Yes | ½ | Scalar | An approximate solution of BWE | Schrödinger equation |

### 9.3. Quantum does not mean "corpuscle-like"

From Planck's study, we know the energy of radiation wave is transmitted in packets ("energy quanta") [8]. This leads to the interpretation that light is a particle (called "photon") [9]. This particle view may give people the impression that the photon is like a corpuscular object. But this is not true at all. The appearance of energy quantum only reflects the *principle of all-or-none* in the emission and absorption of radiation wave. That does not necessarily imply that the photon is a pointed object. From classical electrodynamics, we know light is an electro-magnetic wave. The wave length of visible light is quite long (about 500 nm), which is much larger than the diameters of an atom. Thus the photon cannot be regarded as a pointed object.

This understanding is highly useful for explaining the recent satellite observations of the cosmic microwave background (CMB) [45], [46]. According to the Big Bang theory [47], in the early day of the universe, it is filled with energetic photons. After the universe cooled down, these primordial photons became the CMB that we can detect today [48]. From the recent satellite measurements, the microwave detected in the CMB follows perfectly the Planck's law of black-body radiation [49]. Thus, the microwave radiation must obey $E = h\nu$. Does this imply that the microwave radiation is transmitted in the form of discrete particle?

It is obvious that the microwave is a wave packet of electromagnetic radiation. It cannot be regarded as a pointed object, since its wave length is quite long. Then, how to explain the fact that CMB satisfies the Planck's relation? We think the answer is that a particle does not need to be a pointed object; a wave packet can behave like a particle. Our recent work showed that this is indeed the case [50]. Based on the Maxwell theory, we showed that the Planck's relation and de Broglie relation can be derived by treating the photon as a wave packet of radiation wave. The observation of energy quantization only means that the wave packet has a critical size; the wave packet cannot be arbitrarily small. Its emission and transmission follow the *principle of all-or-none* [50].

Similarly, in the case of electron, although it satisfies both the Planck's relation and the de Broglie relation, it does not mean the electron is a pointed object. As we know from observations in diffraction experiments, the electron clearly has wave properties [20]. It behaves very much like a photon. Thus, it is not unreasonable to regard the electron as an excitation wave of the vacuum.

We may add that, since the particle is a wave packet, its actual wave function is slightly more complicated than the wave function obtained from the conventional quantum wave equations. The solution of the quantum wave equation usually represents a continuous wave, not a wave packet. Such a continuous wave spreads over a very large area and time. The particle, on the other hand, is highly localized. So, in reality, the particle is represented by a wave packet which is made up with overlapping wave functions with slightly different frequencies. For more detailed discussion of this point, please see our recent publication [50].

**9.4. Long-range and short-range forces in the vacuum**

In this work, we proposed that free particles are excitation waves of the vacuum. These free particles include photons, scalar particles, electrons and other leptons. They are stand-alone particles with no internal sub-component. On the other hand, we know there are other sub-atomic particles, like protons and neutrons, which are composed of sub-particles (quarks). These are composite particles called "hadrons". Are these hadrons excitation waves of the vacuum too?

We think the answer is "yes". However, there are basic differences between the elementary particles discussed so far and the hadrons. The size of the photons and electrons are apparently relatively large (in comparison to an atomic nucleus). Their wavelength could be very long, (say, from Å to mm). The hadron's size is very small; the wavelength could be less than $10^{-15}$ m. We think the photon/lepton and the hadron belong to two different types of excitation waves of the vacuum.

It is well known that there are long-range and short-range forces in nature. The electro-magnetic force is a long-range force. In this paper, we propose that this long-range force is associated with the stress/strain of the vacuum medium. The photons and leptons are apparently the excitation waves driven by this long-range force. The short-range forces are known to include strong and weak nuclear forces. We believe they are associated with a different type of physical properties of the vacuum medium. Thus, there could be two different types of excitation waves transmitted through the vacuum: The first type is driven by the long-range force; photons, scalar particles, electrons and other leptons all belong to this type. The second type is driven by the short-range forces. It includes all hadrons.

At present, we have very good knowledge about the physical properties of the long-range force. They are basically described by the Maxwell equations. Thus, we can derive the wave equations for the free particles associated with the excitation waves of the long-range force. For the short-range forces in the vacuum medium, their physical properties are not yet clearly known. Thus, it is very difficult to write down quantum wave equations for the hadrons.

At this point, one can see that our visible world is really made up of different combinations of waves. We know all matters are composed of atoms, each of which in turn is composed of a nucleus (hadrons) and surrounding electrons. Since both hadrons and electrons are excitation waves of the vacuum, the entire atom is basically made up of waves. We may also notice that, the atom by nature is a dynamic object. Even an isolated (or cold-trapped) atom is still not a stationary object.

### 9.5. Origin of matter in our universe

In the study of cosmology, the most important challenge is to explain how matter could be generated at the beginning of our universe. The leading theory today is the Big Bang theory [47]. It is generally believed that matters are essentially converted from energy which is emerged in a fraction of second at the origin of the universe [28], [47]. But how could matter be converted from energy? What is the physical process involved? This paper suggests a simple explanation. Since matters are composed of sub-atomic particles, which are excitation waves of the vacuum, the energy generated by the Big Bang can be used to excite the vacuum to generate a huge number of excitation waves. Since different types of excitation waves represent different types of particles, the Big Bang can produce a great variety of particles at the very early time of the universe. Some of these particles could be short-live, while others could be long-live. Some may have positive charges, while others may have negative charges. When the universe gradually cooled down, the positive charged and negative charged particles can attract each other to form atoms. These atoms eventually form stars, planets, galaxies and even living organisms.

This model, in fact, not only can explain the origin of the visible matters in our universe, it can also provide an explanation to the existence of dark matter. According to the recent study of CMB radiation, our universe appears to contain about 5% ordinary matter, 26% dark matter and 69% dark energy [45], [46]. So, the amount of dark matter is about five times of that of visible matter. At this point, very little is known about the dark matter. No one knows what the dark matter is composed of. The leading hypothesis about dark matter is that it is composed of *weakly interacting massive particles* (WIMPs) which interact only through gravity and the weak nuclear force [51]. So far, none of the experiments designed to detect WIMPs has detected any of such particles [52]-[56].

The matter wave model discussed in this work may offer a useful approach to explain the existence of dark matters. According to our model, all particles in nature are excitation waves of the vacuum. Some of these excitation waves (particles) could behave as dark matter. These particles do not interact easily with other particles because:

a) They have no electric charge. Thus, they do not interact with other charged particles such as electrons.

b) They are not hadrons. Since they are not made of quarks, they cannot interact with the atomic nuclei and thus can easily pass through ordinary matters.

c) They have very small cross sections in interacting with visible matters, just like neutrinos.

d) Unlike the photons, they possess large resting energy. For light to be absorbed by an orbital electron within an atom, it requires a matching of the incoming particle energy with the electron transition energy. For most condensed matters, the energy level of an orbital electron is in the order of several electron volts. If the energy of the incoming particle has rest mass, its total energy is much larger than the electron transition energy. Thus, the chance of absorption is extremely small.

The properties listed above are not unusual. It is reasonable to expect that many excitation waves of the vacuum generated at the Big Bang can satisfy those requirements. Thus, the number of particles meeting these criteria could be very large. Their number could be much larger than the particles forming visible matters. We did not detect them before because they do not interact easily with visible matters.

### 9.6. Advantages of this matter wave model

In science, a major reason for proposing a new model is to provide more simple explanations to experimental facts. Can the matter wave model proposed in this work satisfy such a criterion? We think it can. This model can easily explain many observations that were previously difficult to explain. For example:

(1) It explains why electrons behave very similar to photons. As demonstrated in diffraction experiments and double-slit experiments, we know the electron behaves just like a light wave. And, both electrons and photons are transmitted in discrete energy quanta. Such similarities between electrons and photons had been recognized by many physicists. For example, according to Richard Feynman: "…*electrons behave just like light. The quantum behavior of atomic* *objects ... is the same for all, they are all 'particle waves,'...so what we learn about the properties of electrons... will apply also to all 'particles,' including photons of light*" [44]. The existing theories did not explain the physical basis of such similarities, but our wave model can explain it.

(2) <u>All particles have the same speed limit in traveling through the vacuum</u>. It is well known that no particle can travel faster than the speed of light. But why is it? Previously, there was no explanation to this well-tested experimental fact. It was just regarded as a postulate in STR. With this wave model, it is very easy to explain it. We know the speed of wave propagation is dependent on the physical properties of the transmitting medium [13]. If all particles are excitation waves of the vacuum medium, they should all have the same speed limit.

(3) <u>Massive particles and photons can be converted between each other</u>. As we have discussed earlier, electron-positron pair can be created from energetic photons and vice versa. This may suggest that both electrons and photons are excitation waves of the same vacuum medium. In fact, such conversion is not limited to photons and electrons. In collider physics today, one can frequently observe conversion of particles between different types. For example, a simple collision event is the production of $\mu+/\mu-$ pair by $e+/e-$ collisions. It is generally believed that the intermediate product in this collision is a gamma ray. In such an example, one can see that a photon can be generated by the annihilation of the electron and positron pair. And, the annihilation of a photon can create a pair of muon and anti-muon. The fact that massive particles and photons can be converted between each other strongly suggests that both of them are excitation waves of the same medium.

Another judgement of a model is whether its hypotheses are reasonable. A key assumption of this wave model is that the vacuum behaves like a medium rather than an empty space. This assumption is not unusual. In fact, both the Standard Model of cosmology and the Standard Model of particle physics require a non-empty vacuum, which is regarded as the ground state of the quantum fields [28]-[30]. Furthermore, according to the Maxwell theory of electro-magnetism, the vacuum is like a dielectric medium. Whether the vacuum is an empty space or not is a fundamental question in physics. It will be subject to many tests. Recently, we proposed such an experimental test based on precise measurements of the particle mass in different inertial frames [57]. Hopefully, results of this experiment will be able to settle this key question in the future.

## 10. Conclusion

In this work, we propose that the concept of quantum mechanics can be a natural extension of classical mechanics. Thus, the derivation of the quantum wave equations can be traced to a clear physical mechanism, just like the derivation of classical wave equations. Our approach is to regard *all free particles as excitation waves of the vacuum; different types of particles are represented by different excitation modes*. Thus, both radiation waves (e.g., photons) and matter waves (e.g., electrons) are excitation waves of the same medium.

The major findings of this work are the followings:

- In both the mechanical medium and the vacuum medium, their wave movements are typically described by the *4-D Laplace equation*. This is true for mechanical waves, radiation waves and matter waves.

- The excitation of the vacuum medium can be transmitted as a travelling wave packet. Such a wave packet behaves like a particle since it obeys the Planck's relation and the de Broglie relation. The speed of this traveling wave is entirely determined by the physical properties of the medium. This explains why no particle can travel faster than light.

- In all different types of excitation waves, the wave function always represents the oscillation of a *basic field*, which either directly or indirectly describes the *displacement* of the medium.

- In the case of matter wave, the wave function appears to represent the oscillation of the electric vector potential **Z**, which is related to the electric displacement **D** by the Helmholtz decomposition.

- The motion of the matter wave is commonly described by a *Basic Wave Equation* (BWE). Since all particles are special excitation modes of the matter wave, all known quantum mechanical wave equations are derivatives of the BWE.

- The general solution of BWE involves both a transverse and a longitudinal (*path*) component. The wave function of the Klein-Gordon equation ($\phi$) represents only the *path* component of the matter wave.

- The Dirac equation can also be derived from the BWE. The Dirac wave function ($\psi$) is a special solution of the *path* component of the matter wave.

- The Schrödinger equation can also be derived based on the BWE. We showed that the Schrödinger wave function ($\psi_s$) is a global transformation of the *path* component of the matter wave ($\phi$), i.e., $\psi_s = e^{imc^2 t/\hbar} \phi$.

- This work suggests that, in addition to the statistical meaning proposed by the Copenhagen interpretation, the quantum wave function can also have a physical meaning, namely, it represents a physical motion of the vacuum.

- This model provides a possible explanation for the origin of visible matters and dark matters in our universe.

**Acknowledgements:** I thank Drs. John A. Wheeler, H. E. Rorschach, Bambi Hu, Don Tow, Zhaoqing Zhang, and Xiangrong Wang for providing helpful comments. Particularly, Dr. Yi-Kuen Lee provided valuable inputs on the wave mechanism of an elastic solid. I also thank Ms. Lan Fu for assistance. This work was partially supported by the Research Grant Council of Hong

# Appendix: Vector fields and the Helmholtz decomposition theorem

In many branches of physics, some of the physical parameters can be expressed as *vector fields*. The Helmholtz theorem states that, any sufficiently smooth vector field in a 3-dimensional space can be resolved into the sum of an irrotational (*curl-free*) vector field and a solenoidal (*divergence-free*) vector field, i.e.,

$$\mathbf{F} = -\nabla\Phi + \nabla\times\mathbf{A}, \tag{A1}$$

where $\Phi$ is called the "*scalar potential*", and **A** is called the "*vector potential*". This procedure is known as *Helmholtz decomposition* [16], [58].

In physics, the *curl-free component* of a vector field is often referred to as the "*longitudinal component*" and the divergence-free component is referred to as the "*transverse component*" [59]. Thus, a vector field can be generally written as

$$\mathbf{F} = \mathbf{F}_\ell + \mathbf{F}_t, \tag{A2}$$

where $\mathbf{F}_\ell = -\nabla\Phi$ and $\mathbf{F}_t = \nabla\times\mathbf{A}$. In a 3-dimensional space, both the scalar potential and the vector potential can be directly derived from the vector field **F** [59]. Furthermore, since the choice of **A** is not unique, one can always set $\nabla\cdot\mathbf{A} = 0$ by choosing a specific *gauge* condition. This can be shown as follows:

Suppose $\mathbf{A}'$ is a choice for the vector potential which satisfies Eq. (A1) but $\nabla\cdot\mathbf{A}' \neq 0$. We can redefine a new vector potential **A** such that

$$\mathbf{A} = \mathbf{A}' + \nabla\chi, \tag{A3}$$

where $\chi$ is an arbitrary scalar function. Since

$$\mathbf{F}_t = \nabla\times\mathbf{A}' = \nabla\times(\mathbf{A}-\nabla\chi) = \nabla\times\mathbf{A} - \nabla\times\nabla\chi = \nabla\times\mathbf{A}, \tag{A4}$$

$\mathbf{F}_t$ is unchanged by the above redefinition. Now, since $\chi$ can be any arbitrary function, we can choose $\chi$ in such a way that $\nabla^2\chi = -\nabla\cdot\mathbf{A}'$. Then, we have

$$\nabla\cdot\mathbf{A} = \nabla\cdot(\mathbf{A}'+\nabla\chi) = \nabla\cdot\mathbf{A}' + \nabla\cdot\nabla\chi = 0. \tag{A5}$$

# References

[1] J. C. Slater, *Concepts and Development of Quantum Physics*. Dover Publications, 1969, p 170.
[2] T. Hey and P. Walters, The New Quantum Universe. Cambridge Univ. Press, 2003, p 35.
[3] R. P. Feynman, R. B. Leighton and M. Sands, *Feynman Lectures on Physics*. Reading, Mass: Addison-Wesley, 1964, p 16.12.
[4] P. A. M. Dirac, *The Principles of Quantum Mechanics*. Oxford: Clarendon Press, 1981, p 87.
[5] A. Messiah, *Quantum Mechanics*, Vol. 1. New York: John Wiley & Sons, 1965, pp. 47-48.
[6] A. Messiah, *Quantum Mechanics*. Vol. 1. New York: John Wiley & Sons, 1965, pp. 45-59.
[7] R. P. Feynman, R. B. Leighton and M. Sands, *Feynman Lectures on Physics*. Reading, Mass: Addison-Wesley, 1964, pp. 1.1-1.8.
[8] M. Planck, "Zur theorie des gesetzes der energieverteilung im normalspectrum," *Verhandl. Dtsch. Phys. Ges.*, **2**, pp. 237-245, 1900.
[9] A. Einstein, "Concerning an heuristic point of view toward the emission and transformation of light," *Am. J. Phys.*, **33**, 367, 1965.
[10] A. Messiah, *Quantum Mechanics*, Vol. 1. New York: John Wiley & Sons, 1965, pp. 27-42.
[11] H. Reismann and P. S. Pawlik, *Elasticity, Theory and Applications*. Wiley, 1980.
[12] K. F. Graff, *Wave Motion in Elastic Solids*. Clarendon Press, 1975 (and Dover Publications, 1991).
[13] D. C. Chang and Y. Lee, "Study on the physical basis of wave-particle duality: Modelling the vacuum as a continuous mechanical medium," *J. Mod. Phys.*, **6**, pp. 1058-1070, 2015.
[14] Y. C. Fung, *A First Course in Continuum Mechanics*. NJ: Prentice-Hall, 1977.
[15] J. Salencon, *Handbook of Continuum Mechanics: General Concepts, Thermoelasticity*. Springer, 2001, p 333.
[16] G. B. Arfken and H. J. Weber, *Mathematical Methods for Physicists*. Cambridge Univ. Press, 1995.
[17] A. Messiah, *Quantum Mechanics*, Vol. 1. New York: John Wiley & Sons, 1965, pp. 45-74.
[18] J. C. Slater, *Concepts and Development of Quantum Physics*. Dover Publications, 1969, p 177.
[19] G. P. Thomson and A. Reid, "Diffraction of Cathode Rays by a thin Film," *Nature*, **119**, p 890, 1927.
[20] C. J. Davisson and L. H. Germer, "The scattering of electrons by a single crystal of nickel," *Nature*, **119**, pp. 558–560, 1927.
[21] R. Bach et al, "Controlled double-slit electron diffraction," *New J. Phys.*, **15**, 033018, 2013.
[22] R. F. Egerton, *Physical Principles of Electron Microscopy*. Springer, 2016.
[23] P. A. M. Dirac, *The Principles of Quantum Mechanics*, 4th ed., Oxford: Clarendon Press, 1981, pp. 253-275.
[24] J. J. Sakurai, *Advanced Quantum Mechanics*. Reading: Addison-Wesley, 1973, p 132.
[25] E. Whittaker, *A History of the Theories of Aether and Electricity*. London: Thomas Nelson and Sons, 1951.
[26] A. A. Michelson and E. W. Morley, "On the relative motion of the Earth and the luminiferous ether," *Am. J. Sci.*, **34**, 1887, pp. 333-345.
[27] A. Einstein, "Zur elektrodynamik bewegter körper," *Ann. Phys.*, **322**, pp. 891-921, 1905.
[28] A. H. Guth and D. I. Kaiser, "Inflationary cosmology: Exploring the universe from the smallest to the largest scales," *Science*, **307**, pp. 884-890, 2005.
[29] A. Messiah, *Quantum Mechanics*, Vol. 1. New York: John Wiley & Sons, 1965, p 439.
[30] L. H. Ryder, *Quantum Field Theory*. Cambridge; New York: Cambridge Univ. Press, 1996, p 284.

[31] D. C. Chang, "A classical approach to the modeling of quantum mass," *J. Mod. Phys.*, **4**, pp. 21-30, 2013.
[32] D. C. Chang, "Why energy and mass can be converted between each other? A new perspective based on a matter wave model," *J. Mod. Phys.*, **7**, pp. 395-403, 2016.
[33] M. S. Longair, *Theoretical Concepts in Physics.*1st ed. Cambridge; New York: Cambridge Univ. Press, 1984, pp. 37-59.
[34] W. N. Cottingham and D. A. Greenwood, *An Introduction to the Standard Model of Particle Physics*. Cambridge: Cambridge Univ. Press, 1998, pp. 37-47.
[35] R. P. Feynman, R. B. Leighton and M. Sands, *Feynman Lectures on Physics*. Reading, Mass: Addison-Wesley, 1964, pp. 15.1-15.15.
[36] D. C. Chang, "What is rest mass in the wave-particle duality? A proposed model," *arXiv Preprint Physics/0404044*, 2004.
[37] P. S. Faragó and L. Jánossy, "Review of the experimental evidence for the law of variation of the electron mass with velocity," *Nuovo Cim*, **5**, 1411, 1957.
[38] A. Einstein, *Relativity: The Special and the General Theory*. New York: Three River Press, 1961, pp. 1-64. See also, A.P. French, *Special Relativity* (Nelsen, London, 1968).
[39] R. Shankland, *Atomic and Nuclear Physics*. New York: MacMillan, 1961, p 207.
[40] A. Messiah, *Quantum Mechanics*. Vol 1, New York: John Wiley & Sons, 1965, pp. 63-65.
[41] J. J. Sakurai, *Advanced Quantum Mechanics*. Reading: Addison-Wesley, 1973, pp. 75-89.
[42] W. N. Cottingham and D. A. Greenwood, *An Introduction to the Standard Model of Particle Physics*. Cambridge: Cambridge Univ. Press, 1998, pp. 71-72.
[43] T. Hey and P. Walters, *The New Quantum Universe*. Cambridge Univ. Press, 2003, p 15.
[44] R. P. Feynman, R. B. Leighton and M. Sands, *Feynman Lectures on Physics*. Reading, Mass: Addison-Wesley, 1964, pp. 37.1-37.12.
[45] C. L. Bennett and et.al., "Nine-Year Wilkinson Microwave Anisotropy Probe (WMAP) Observations: Final Maps and Results," *Astrophys. J. Supplement Series*, **208**, 20B, 2013.
[46] Planck Collaboration, P. A. R. Ade and et al, "Planck 2015 results - XIII. Cosmological parameters," *Astron. Astrophys.*, **594**, A13, 2016.
[47] P. Coles, "The state of the Universe," *Nature*, **433**, pp. 248-256, 2005.
[48] R. H. Dicke et al, "Cosmic Black-Body Radiation," *Astrophys. J.*, **142**, pp. 414-419, 1965.
[49] J. C. Mather et al, "Measurement of the cosmic microwave background spectrum by the COBE FIRAS instrument," *Astrophys. J.*, **420**, pp. 439-444, 1994.
[50] D. C. Chang, "Physical interpretation of the Planck's constant based on the Maxwell theory," *Chin. Phys. B*, 2017 in press.
[51] G. Bertone, "The moment of truth for WIMP dark matter," *Nature*, **468**, pp. 389-393, 2010.
[52] D. N. Spergel, "The dark side of cosmology: dark matter and dark energy," *Science*, **347**, pp. 1100-1102, 2015.
[53] LUX Collaboration, D. Akerib and et al, "First results from the LUX dark matter experiment at the Sanford Underground Research Facility," *Phys. Rev. Lett.*, **112**, 091303, 2014.
[54] XENON100 Collaboration, E. Aprile and et al, "Limits on spin-dependent WIMP-nucleon cross sections from 225 live days of XENON100 data," *Phys. Rev. Lett.*, **111**, 021301, 2013.
[55] The CDMS II Collaboration, "Dark Matter Search Results from the CDMS II Experiment," *Science*, **327**, pp. 1619-1621, 2010.

[56] SuperCDMS Collaboration, R. Agnese and et al, "Search for Low-Mass Weakly Interacting Massive Particles with SuperCDMS," *Phys. Rev. Lett.*, **112**, 241302, 2014.

[57] D. C. Chang, "Is there a resting frame in the universe? A proposed experimental test based on a precise measurement of particle mass," *Euro. Phys. J. Plus*, 2017 in press.

[58] H. Helmholtz, "Über Integrale der hydrodynamischen Gleichungen, welcher der Wirbelbewegungen entsprechen," *J. Reine Angew. Math.*, **55**, pp. 25-55, 1858.

[59] J. D. Jackson, *Classical Electrodynamics*. New York: Wiley, 1962, pp. 178-199